\documentclass[fleqn,usenatbib]{mnras}

\usepackage{newtxtext,newtxmath}

\usepackage[T1]{fontenc}

\DeclareRobustCommand{\VAN}[3]{#2}
\let\VANthebibliography\thebibliography
\def\thebibliography{\DeclareRobustCommand{\VAN}[3]{##3}\VANthebibliography}

\usepackage[export]{adjustbox}
\usepackage{lineno}

\usepackage{graphicx}	
\usepackage{amsmath}	
\usepackage{amssymb}	

\title[ZTF SN Siblings]{Supernova Siblings and their Parent Galaxies in the Zwicky Transient Facility Bright Transient Survey}

\author[M. L. Graham al.]{Melissa L. Graham$^{1}$\thanks{E-mail: mlg3k@uw.edu},
Christoffer Fremling$^{2}$,
Daniel A. Perley$^{3}$,
Rahul Biswas$^{4}$,
Christopher A. Phillips$^{1}$,
\newauthor
Jesper Sollerman$^{5}$, 
Peter E. Nugent$^{6,7}$, 
Sarafina Nance$^{7,6}$, 
Suhail Dhawan$^{8}$,
Jakob Nordin$^{9}$,
Ariel Goobar$^{4}$,
\newauthor
Adam Miller$^{10,11}$,
James D. Neill$^{2}$, 
Xander J. Hall$^{2}$,
Matthew J. Hankins$^{12}$,
Dmitry A. Duev$^{2}$, 
\newauthor
Mansi M. Kasliwal$^{2}$, 
Mickael Rigault$^{13}$, 
Eric C. Bellm$^{1}$, 
David Hale$^{14}$,
Przemek Mr\'{o}z$^{2}$,
and S.~R.~Kulkarni$^{2}$
\\
$^{1}$DIRAC Institute, Department of Astronomy, University of Washington, 3910 15th Avenue NE, Seattle, WA 98195, USA \\
$^{2}$Division of Physics, Mathematics, and Astronomy, California Institute of Technology, Pasadena, CA 91125, USA \\
$^{3}$Astrophysics Research Institute, Liverpool John Moores University, IC2, Liverpool Science Park, 146 Brownlow Hill, Liverpool L3 5RF, UK \\
$^{4}$The Oskar Klein Centre, Department of Physics,  Stockholm University, SE-106 91 Stockholm , Sweden \\
$^{5}$The Oskar Klein Centre, Department of Astronomy,  Stockholm University, SE-106 91 Stockholm , Sweden \\
$^{6}$E.O. Lawrence Berkeley National Laboratory, 1 Cyclotron Rd., Berkeley, CA, 94720 \\
$^{7}$Department of Astronomy, University of California, Berkeley, CA 94720-3411, USA \\
$^{8}$Institute of Astronomy and Kavli Institute for Cosmology, University of Cambridge, Madingley Road, Cambridge CB3 0HA, UK \\
$^{9}$Institute of Physics, Humboldt-Universit\"{a}t zu Berlin, Newtonstr. 15, 12489 Berlin, Germany \\
$^{10}$Center for Interdisciplinary Exploration and Research in Astrophysics and Department of Physics and Astronomy, Northwestern University, \\ 1800 Sherman Ave, Evanston, IL 60201, USA \\
$^{11}$The Adler Planetarium, Chicago, IL 60605, USA \\
$^{12}$Arkansas Tech University Department of Physical Sciences, 1701 N. Boulder Avenue, Russellville, AR 72801 USA \\
$^{13}$Univ Lyon, Univ Claude Bernard Lyon 1, CNRS, IP2I Lyon / IN2P3, IMR 5822, F-69622, Villeurbanne, France \\
$^{14}$Caltech Optical Observatories, California Institute of Technology, Pasadena, CA  91125
}

\date{Accepted XXX. Received YYY; in original form ZZZ}

\pubyear{2015}

\begin{document}
\label{firstpage}
\pagerange{\pageref{firstpage}--\pageref{lastpage}}
\maketitle

\begin{abstract}
Supernova (SN) siblings -- two or more SNe in the same parent galaxy -- are useful tools for exploring progenitor stellar populations as well as properties of the host galaxies such as distance, star formation rate, dust extinction, and metallicity.
Since the average SN rate for a Milky Way-type galaxy is just one per century, a large imaging survey is required to discover an appreciable sample of SN siblings.
From the wide-field Zwicky Transient Facility (ZTF) Bright Transient Survey (BTS; which aims for spectroscopic completeness for all transients which peak brighter than $r{<}$18.5 mag) we present 10 SN siblings in 5 parent galaxies.
For each of these families we analyze the SN's location within the host and its underlying stellar population, finding agreement with expectations that SNe from more massive progenitors are found nearer to their host core and in regions of more active star formation.
We also present an analysis of the relative rates of core collapse and thermonuclear SN siblings, finding a significantly lower ratio than past SN sibling samples due to the unbiased nature of the ZTF.
\end{abstract}







\section{Introduction}\label{sec:intro}

It has long been known that some galaxies are more efficient at producing supernovae than others.
In describing their choice of fields to survey in order to generate a large sample of supernovae, \citet{1938ApJ....88..529Z} focus on galaxies in nearby clusters (i.e., where a single field of view can contain a large stellar mass), and on star-forming galaxies such as those similar to the Andromeda Galaxy (grand design spirals) and those with low-surface brightness, which offer the additional bonus of being easy to search by eye.
Such a focus is effective, especially when survey \'{e}tendue\footnote{The product of the camera's field of view and the telescope's aperture.} is limited, however this strategy can miss entire populations of transient events (i.e., those which are hostless or in galaxies of low stellar mass), and does not work well for high-redshift cosmological applications.

The Zwicky Transient Facility (ZTF; \citealt{2019PASP..131g8001G,2019PASP..131a8002B,2019PASP..131a8003M}) is a modern wide-field optical sky survey which does not need to target individual galaxies thanks to its 47 deg$^2$ field of view. 
The ZTF is a public-private partnership survey which uses the 48 inch telescope at Palomar Observatory to image the entire northern sky once every $\sim3$ days in the $g$ and $r$ filters to a depth of $r\sim20.5$ mag \citep{2019PASP..131f8003B}. 
ZTF images are processed in real-time using the difference imaging algorithm of \citet{2016ApJ...830...27Z}.
Discoveries made in the public data are released as alerts \citep{2019PASP..131a8001P}, and available via the public alert brokers such as 
AMPEL\footnote{\citet{2019A&A...631A.147N}; \url{https://ampelproject.github.io/}},  
ANTARES\footnote{\citet{2021AJ....161..107M}; \url{https://antares.noirlab.edu/}}, 
ALeRCE\footnote{\citet{2020arXiv200803303F}; \url{http://alerce.science/}}, and 
Lasair\footnote{\citet{2019RNAAS...3...26S}; \url{https://lasair.roe.ac.uk/}}.
Dedicated time on the Palomar Observatory 60 and 200 inch telescopes, the former with the SED Machine instrument \citep[SEDM;][]{2018PASP..130c5003B,2019A&A...627A.115R}, is used for follow-up and classification of ZTF discoveries.  
In particular, the ZTF Bright Transient Survey (BTS; \citealt{2020ApJ...895...32F,2020ApJ...904...35P}) applies a filter to the ZTF public survey alert stream to identify and spectroscopically classify, with very high completeness, transients with a peak apparent brightness of $r\leq18.5$ mag (hereafter referred to as the BTS filter). 
The BTS follow-up strategy is to target all likely SNe detected by ZTF that are brighter than 19 mag in the $g$ or $r$ filter, and then raising the priority for targets brighter than 18.5 mag, with the goal of obtaining spectra for all objects brighter than 18.5 (unless prevented by, e.g., bad weather or technical issues). 
As described in \citet{2020ApJ...904...35P}, this strategy lead to a spectroscopic completeness rate of 93\% (75\%) for SNe brighter than 18.5 (19) mag.

In such large modern surveys it remains true that some galaxies appear to be more prolific producers of SNe, through a combination of galaxy properties and chance. 
SNe which occur in the same host galaxy are referred to as "siblings", and their common host the "parent" galaxy.
These SN siblings are more than just a novelty -- they provide unique scientific opportunities. 
\citet{2009ApJ...698.1307T} study three SN siblings all classified as Type Ib (1999eh, 2007uy, 2008D) in NGC 2770, a spiral galaxy similar to the Milky Way.
They use spectra of the parent galaxy to show that the sites of the SNe\,Ib had subsolar metallicities (but that random chance could not be ruled out as the reason why all three were of the rare SN\,Ib subtype). \citet{2013A&A...550A..69A} compile and analyze a large sample of SN siblings and use the ratio of different types, such as core-collapse supernovae (CC\,SNe; explosions of $>$8 $\rm M_{\odot}$ stars) and Type Ia supernovae (SNe\,Ia, the thermonuclear explosions of white dwarf stars) to constrain the average duration of star-formation episodes.
We discuss their work in more detail in Section~\ref{sec:rates}.

Supernovae are used as cosmological probes, and siblings in the same parent galaxy can be used to study the systematic contribution in their distance estimates. 
\citet{2018A&A...611A..58G} show that the fast-declining SN\,Ia siblings 2007on and 2011iv yield distance estimates to NGC 1404 that are, surprisingly, discrepant by up to $14\%$.
\citet{2020ApJ...895..118B} find that the slow-declining SN\,Ia siblings 2013aa and 2017cbv are nearly identical in their light curves and spectra (although 13aa was not discovered early enough to detect a "blue bump" like 17cbv; \citealt{2017ApJ...845L..11H}).
They also provide a list of SN\,Ia siblings identified in the literature (their Table 3) and show that host-galaxy distances from sibling SNe\,Ia are consistent to within $3\%$ ({\it vs.} $6\%$ among non-siblings).
\citet{2020ApJ...896L..13S} used 16 SN\,Ia siblings from the Dark Energy Survey to show that up to half of the intrinsic scatter in SN\,Ia peak brightness could be attributed to host galaxy properties -- and thus potentially be corrected for in cosmological applications.

In this work we search the ZTF BTS sample for SN siblings and find five pairs, which are presented in Section~\ref{sec:sample}.
We provide individual analyses of these pairs and their host galaxies in Section~\ref{sec:sibs}, examining the SN locations within their hosts, the underlying stellar populations, and in some cases comparing the distance estimates that we have derived from the SN light curves.
In Section~\ref{sec:rates} we provide a relative rates analysis for different sibling types, and discuss how the results from an unbiased survey like ZTF compare with past studies. 
We summarize our conclusions in Section~\ref{sec:conc}.

\section{SN Sibling Identification and Observations}\label{sec:sample}

For this work we begin with all ZTF transients that passed the BTS filter as of Sep 13 2020 (with the very earliest passing on Apr 22 2018), which includes all events -- classified and unclassified -- that were, at any point in their light curve, brighter than $r=19$th magnitude.
This sample was generated using the BTS Explorer webpage \citep{2020ApJ...904...35P}\footnote{\url{https://sites.astro.caltech.edu/ztf/bts/explorer.php}}, with the quality and purity cuts applied.
The quality filter removes candidates which occurred at times or in regions of poor observability, and the purity filter removes objects that are highly likely to be false positives (non-supernova transients, like AGN or variable stars) by requiring that a BTS transient be cross-matched with a galaxy (but not in its nucleus) or have a light-curve with an SN-like timescale (described further in Section 2.4 of \citealt{2020ApJ...904...35P}).
We allow our sample to extend one month earlier and one month later than the start and end dates imposed by \citet{2020ApJ...904...35P}.
Thus we start with a total of $2640$ objects, $148$ more than the total of all transients and unclassified objects quoted in Table 1 of \citet{2020ApJ...904...35P}.
Most of the additional candidates are SNe\,Ia and unclassified transients, the two most populous categories.
In this sample we identified 46 transients that were within $250$\arcsec\ of each other, and another 4 transients with a redshift limit of $z<0.0050$ that were within $250$--$600$\arcsec\ of each other.
All 4 events in the latter group were M31 novae, as were 4 in the previous group, for a total of 8 M31 novae, and 42 potential siblings.

We used the GROWTH Marshal \citep{2019PASP..131c8003K} to visually review images of these 42 potentially associated transients (21 pairs) to identify true siblings, and the results of this visual review are presented in Table~\ref{tab:sib_vis_rev}.
In the top section of Table~\ref{tab:sib_vis_rev} we list the 20 SNe (10 pairs) that we confirm are siblings because they appear in the same parent galaxy and not, e.g., in two close but different host galaxies.
In the middle section, we list 6 SNe (3 pairs) that do not appear in the same host galaxy, and are not siblings, but which might be "cousins" (i.e., their hosts belong to the same group or cluster of galaxies, or their progenitor system might be an intragroup or intracluster star).
In the bottom section of Table~\ref{tab:sib_vis_rev} we list 16 transients (8 pairs) that are within $250$\arcsec\ of each other but do not appear to be physically associated (i.e., are chance alignments).
The 8 novae in M31 which met the BTS criteria are not listed in Table~\ref{tab:sib_vis_rev}\footnote{They are: ZTF19abirmkt, ZTF19abtrjqg, ZTF19acbzgog, ZTF19acgfhfd, ZTF19acnfsij, ZTF19acqprad, ZTF19adakuos, and ZTF20abqhsxb.}.

\begin{table*}
\centering
\caption{The results of a visual review of sibling candidates, sorted by on-sky separation distance and listed in three categories: (top) SNe that we identified as siblings, (middle) SNe whose hosts might be physically associated, and (bottom) chance alignments. In Column 4, the last four characters of a SN's ZTF name are used to refer to the SN where more detail can be offered regarding potential associations between non-sibling SNe.}
\label{tab:sib_vis_rev}
\begin{tabular}{clll}
\hline
Separation ($\arcsec$) & \multicolumn{2}{c}{ZTF Names} & Comments regarding visual review. \\
\hline
\multicolumn{4}{l}{\textit{Identified siblings:}} \\
3.7  & ZTF18aasdted & ZTF19abqhobb & same parent galaxy \\
5.6  & ZTF19aaeiowr & ZTF20aambbfn & same parent galaxy \\
5.6  & ZTF19aaksrgj & ZTF20aavpwxl & same parent galaxy \\
5.7  & ZTF19abgrchq & ZTF19acgaxei & same parent galaxy \\
8.6  & ZTF19accobqx & ZTF19acnwelq & same parent galaxy \\
19.8 & ZTF19aatesgp & ZTF20aaelulu & same parent galaxy \\
23.8 & ZTF19aamhgwm & ZTF19aaqdkrm & same parent galaxy \\
24.3 & ZTF18aboabxv & ZTF18adachwf & same parent galaxy \\
35.7 & ZTF18abdffeo & ZTF19abajxet & same parent galaxy \\
42.4 & ZTF19aavitlq & ZTF19abpyqog & same parent galaxy \\
\hline
\multicolumn{4}{l}{\textit{Not siblings, but hosts might be associated:}} \\
50.1  & ZTF19aaekvwv & ZTF19acnqsui & \emph{qsui} might be a hostless SN in a galaxy cluster with \emph{kvwv}'s host \\
133.8 & ZTF18abdbysy & ZTF20aaurjzv & \emph{rjzv} host might be a satellite galaxy of \emph{bysy} host \\
\hline
\multicolumn{4}{l}{\textit{Not siblings, these appear to be random line-of-sight alignments:}} \\
46.8  & ZTF18abqkfvr & ZTF19aanuipj & \emph{kfvr} appears hostless, might be high-$z$ (unclassified) \\
70.4  & ZTF19aauxmqj & ZTF19abeloei & different hosts; hosts appear to be unassociated (i.e. different $z$) \\
106.1 & ZTF19acykqyr & ZTF20abiserv & \emph{kqyr} appears hostless, is distant from a group hosting \emph{serv} \\
142.1 & ZTF19aavhypb & ZTF18aaizerg & different hosts; hosts appear to be unassociated (i.e. no obvious cluster) \\ 
149.6 & ZTF19aafmymc & ZTF20aazstdx & different hosts; hosts appear to be unassociated (i.e. different $z$) \\
202.8 & ZTF18acxgoki & ZTF19abqgtqo & different hosts; hosts appear to be unassociated (i.e. no obvious cluster) \\
224.7 & ZTF19aarflsx & ZTF20aahggbm & different hosts; hosts appear to be unassociated (i.e. no obvious cluster) \\
237.0 & ZTF19aaeoqst & ZTF19aafndoy & different hosts; might be in the same cluster \\
246.4 & ZTF19acjndrx & ZTF19acjndsa & both are apparently hostless (might be high-$z$) \\
\hline
\end{tabular}
\end{table*}

\begin{table*}
\centering
\caption{Properties of the 20 SNe we identified as siblings. The horizontal line separates the SN siblings for which the brightest observed apparent magnitudes of both events were brighter/fainter than $18.5$ mag.}
\label{tab:sib_cands}
\begin{tabular}{llllllllll}
\hline
 ZTF & IAU &    Spectral & Redshift     & Brightest  & ZTF & IAU & Spectral & Redshift     & Brightest \\
Name & Name &        Type & (Host or SN) & Magnitude* & Name & Name &  Type    & (Host or SN) & Magnitude* \\
\hline
19aatesgp &  2019ehk &       SNIIb & $0.0055$ &  $r=15.82$ & 20aaelulu &   2020oi &        SNIc &    $0.0052$ & $r=13.86$  \\
18aasdted &  2018big &        SNIa & $0.0181$ &  $r=15.72$ & 19abqhobb &  2019nvm &        SNIIP &    $0.0181$ & $r=17.12$  \\
18abdffeo &  2018dbg &      SNIb/c & $0.0148$ &  $r=17.52$ & 19abajxet &  2019hyk &        SNIIP &    $0.0147$ & $g=16.41$  \\
19aamhgwm &  2019bvs &        SNIIL & $0.0342$ &  $r=18.08$ & 19aaqdkrm &  2019dod &        SNIIP &    $0.0342$ & $g=17.98$  \\
19aaeiowr &  2019abo & {\it SNI?} & $0.0432$ &  $g=18.29$ & 20aambbfn &  2020bzv &        SNIa &    $0.0439$ & $r=18.30$  \\
\hline
19abgrchq &  2019lsk &       SNIIb & $0.0300$ & $r=18.16$  & 19acgaxei &  2019svq &           - &    $0.0297$ & $r=18.97$* \\
19accobqx &  2019sik &        SNIa & $0.1000$ & $g=18.52$* & 19acnwelq &  2019uej & {\it SNIa?} & {\it 0.12?} & $g=18.65$* \\
19aavitlq &  2019gip &   SNIa-91bg & $0.0315$ & $r=18.52$* & 19abpyqog &  2019oba &        SNII &    $0.0310$ & $r=18.85$* \\
18aboabxv &  2018fob &        SNIc & $0.0290$ & $r=18.64$* & 18adachwf &  2018lev &       SNIIP &    $0.0290$ & $r=18.84$* \\
19aaksrgj &  2019bbd &        SNIa & $0.0859$ & $g=18.73$* & 20aavpwxl &  2020hzk & {\it SNIa?} &    $0.0859$ & $r=18.82$* \\
\hline
\multicolumn{8}{l}{*Asterisks indicate brightest magnitude is fainter than 18.5 mag.} \\
\end{tabular}
\end{table*}

For our 20 identified SN siblings which passed the BTS filter (10 pairs), we list their SN type, redshift, and brightest observed magnitude in Table~\ref{tab:sib_cands}.
All of this information was obtained from the GROWTH Marshal.
Of these $20$ objects, nine have peak observed brightnesses fainter than $18.5$ mag (denoted by asterisks in columns four and eight).
The $5$ sibling pairs in the BTS sample of SNe with a peak brightness of $<$18.5 mag are all listed above the horizontal line in Table~\ref{tab:sib_cands}.

In the BTS sample of SNe with peak brightness $<$18.5 ($<$19) mag, we found that 9 (16) out of 10 (20) SN siblings were spectroscopically classified.
These fractions match the overall spectral completeness for the full BTS sample of SNe: $75\%$ ($93\%$) for SNe with peak brightness $<$19 ($<$18.5) mag \citep{2020ApJ...904...35P}.

\subsection{Gold Sample: SN Siblings with Peak Brightness $<$18.5 mag}\label{ssec:sample_gold}

The high spectroscopic completeness (93\%) of BTS SNe with peak brightness $<$18.5 mag makes the sample of $5$ sibling pairs with peaks $<$18.5 mag especially useful (e.g., for a rates analysis; Section \ref{sec:rates}).
Hereafter we refer to this set of $5$ sibling pairs as our "Gold Sample" of SN siblings, and focus this paper's analysis on them.

\begin{figure*}
\includegraphics[width=7cm,fbox]{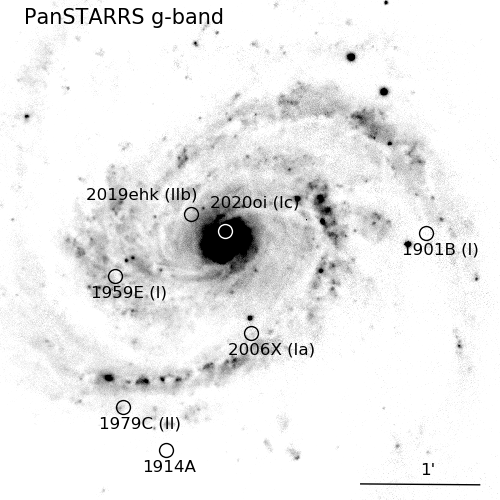}
\includegraphics[width=7cm,fbox]{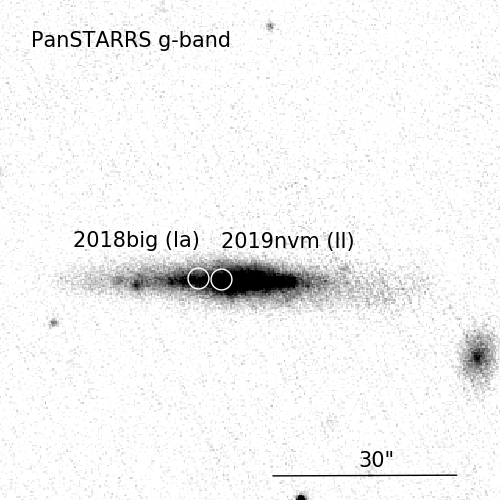}
\includegraphics[width=7cm,fbox]{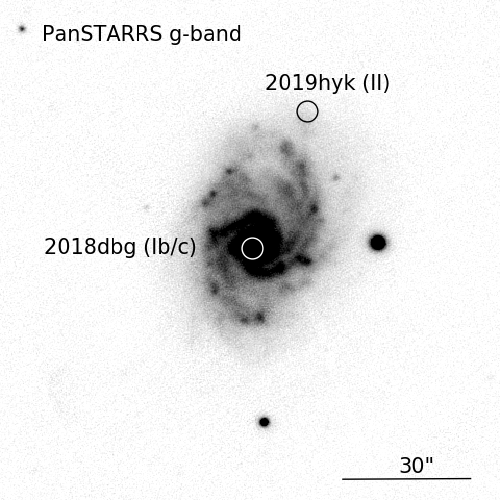}
\includegraphics[width=7cm,fbox]{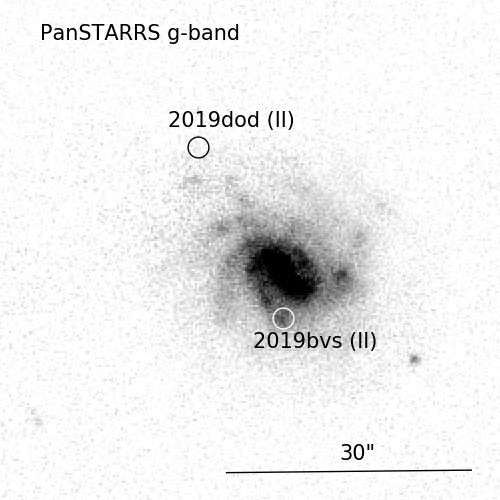}
\includegraphics[width=7cm,fbox]{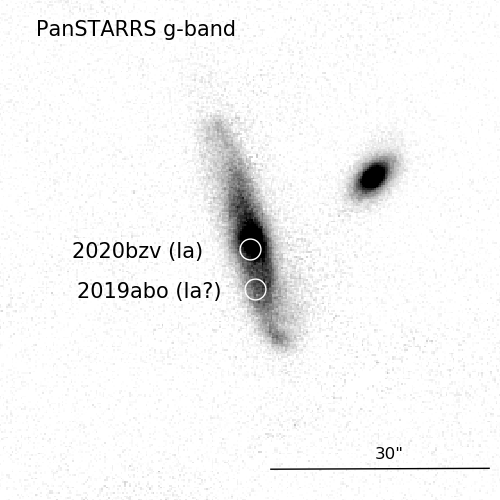}
\caption{Image stamps showing the locations of the five SN sibling pairs with peak observed brightnesses $<$18.5 mag (the Gold Sample). All are north-up, east-left $g$ band images from the PanSTARRS image cutout server, with a linear scaling set to emphasize the host features. Circles mark the locations of SN siblings, as labeled. SN classifications are included in the labels where possible.}
\label{fig:stamps}
\end{figure*}

\begin{figure*}
\includegraphics[width=8cm]{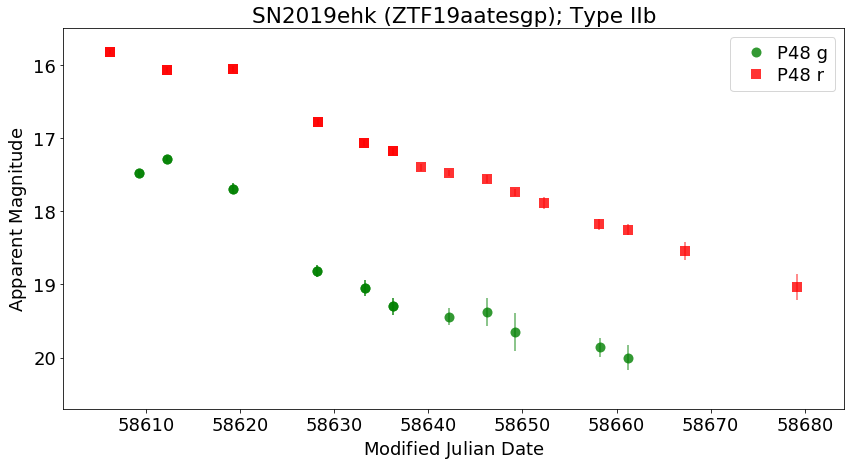}
\includegraphics[width=8cm]{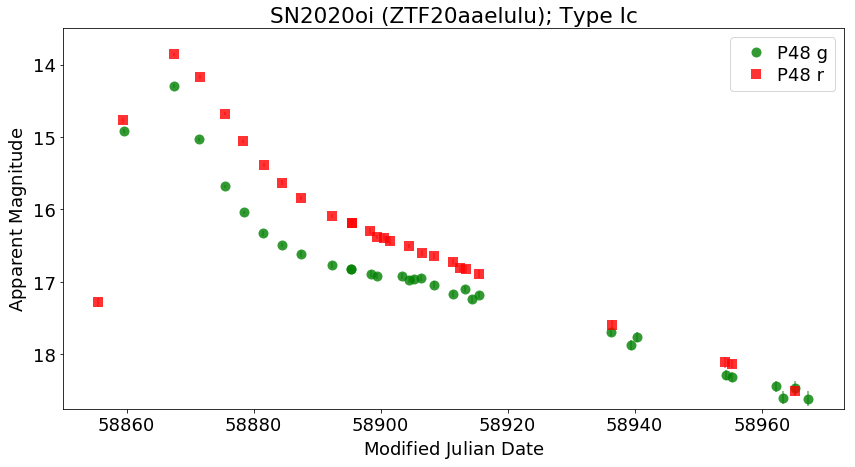}
\includegraphics[width=8cm]{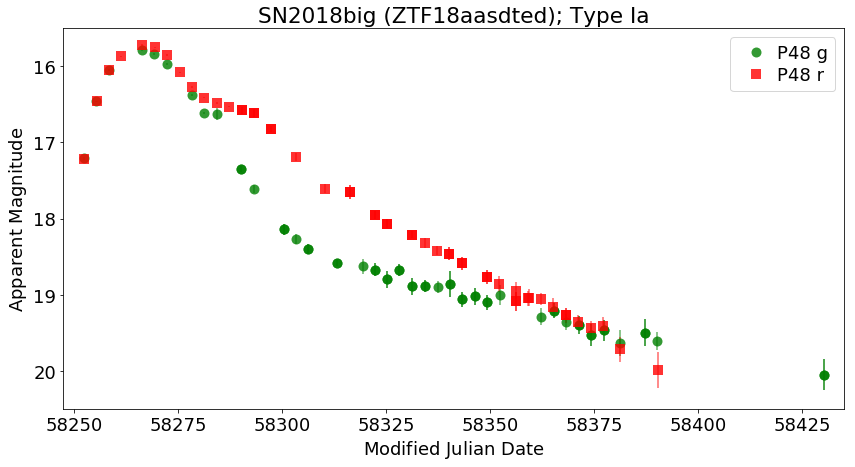}
\includegraphics[width=8cm]{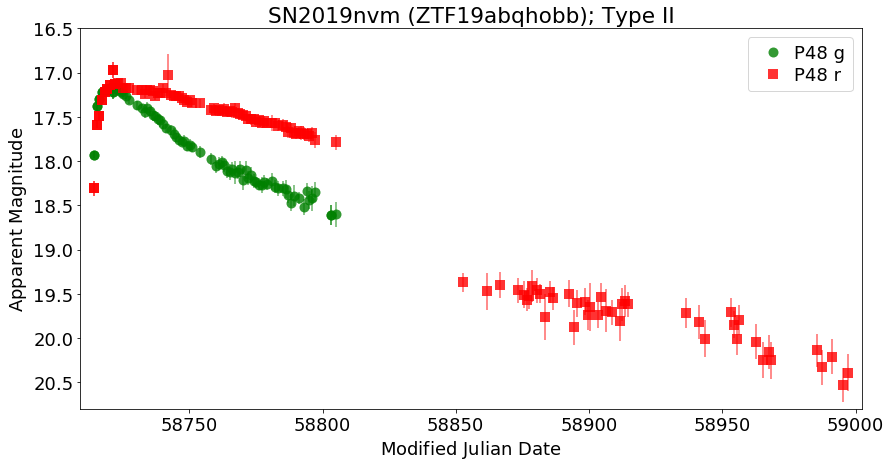}
\includegraphics[width=8cm]{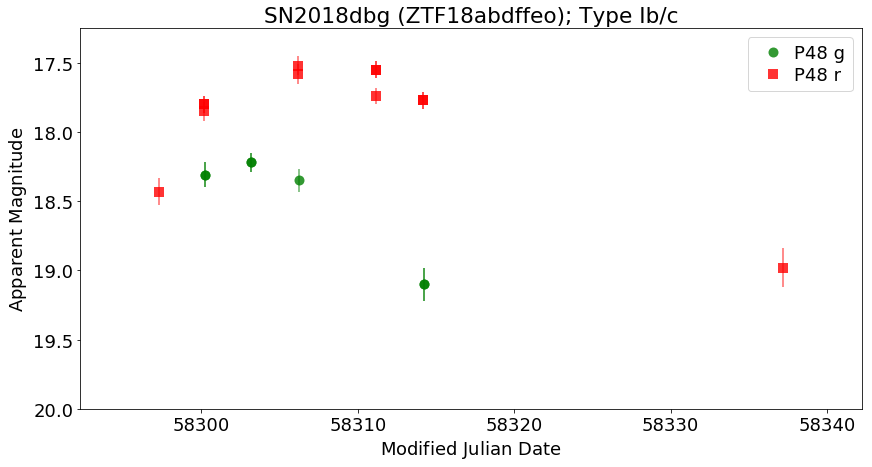}
\includegraphics[width=8cm]{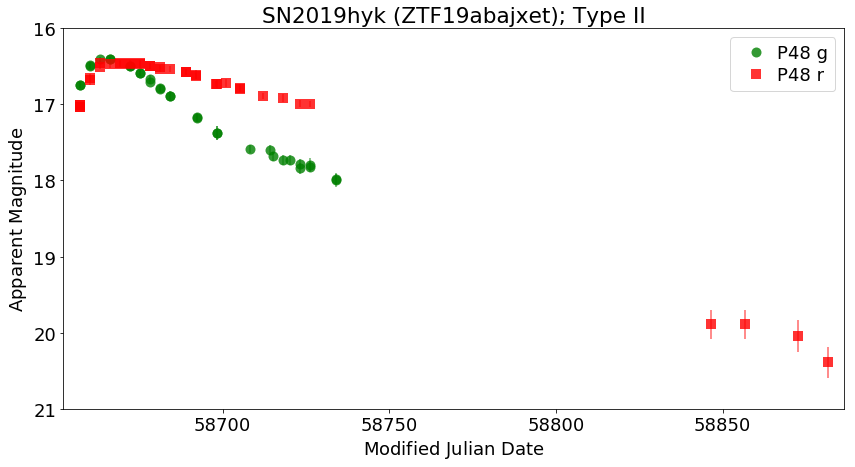}
\includegraphics[width=8cm]{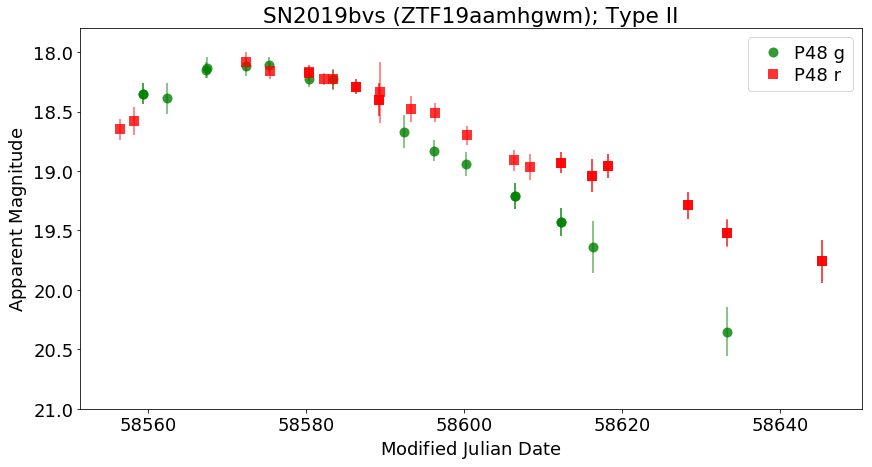}
\includegraphics[width=8cm]{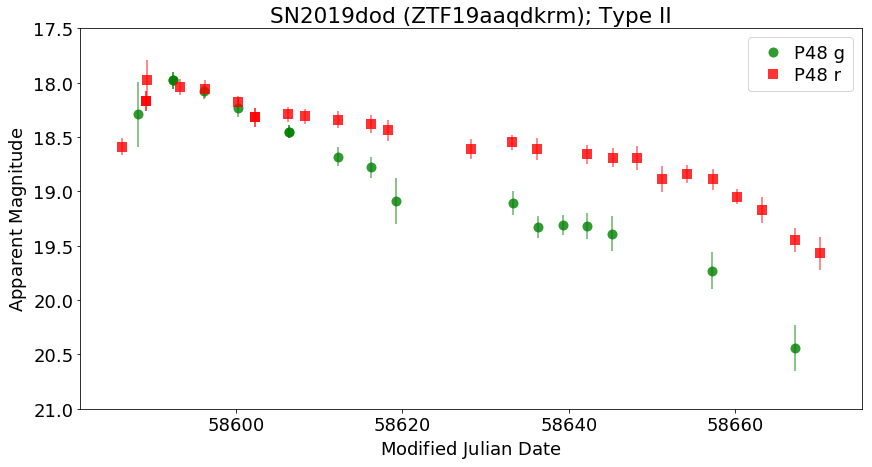}
\includegraphics[width=8cm]{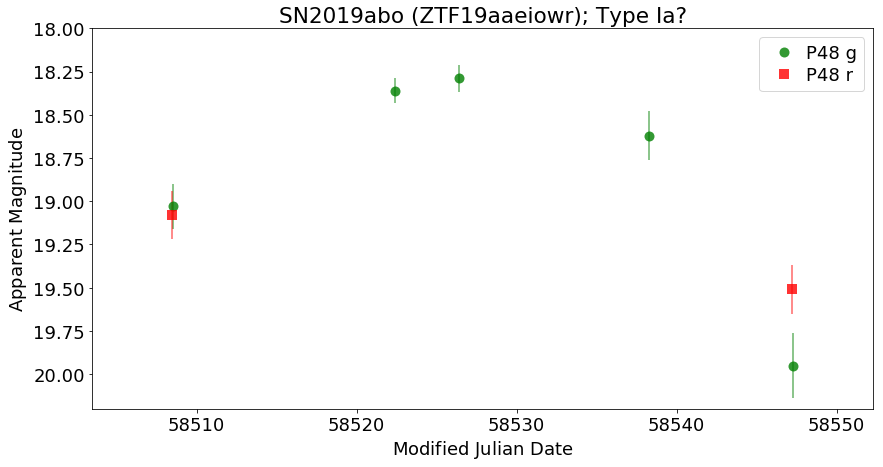}
\includegraphics[width=8cm]{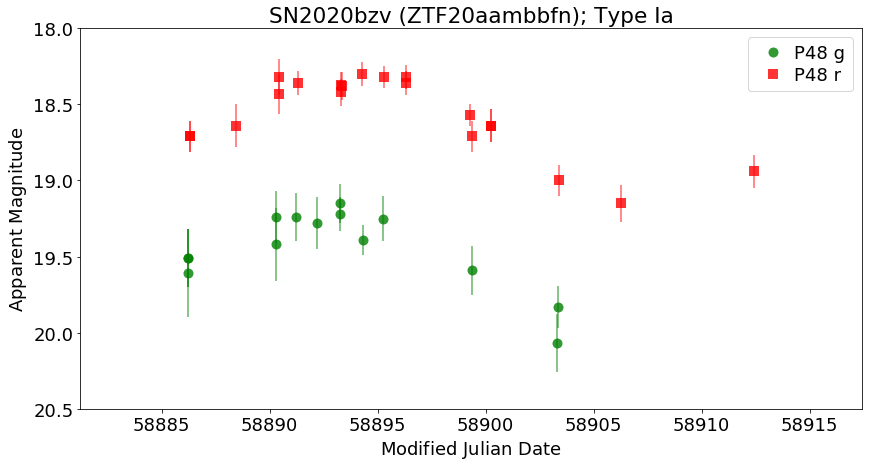}
\caption{Light curves of SN siblings (one pair per row) with peak observed magnitude $\leq18.5$ mag in either the $g$ or $r$ filter (the Gold Sample). Plots show ZTF P48 photometry in \textit{g} and \textit{r} bands (green and red points respectively) from the public data set. Photometry for SN\,2019ehk was published by \citet{2020arXiv200902347D}, and for SN\,2020oi by \citet{2020ApJ...903..132H}. }
\label{fig:phot}
\end{figure*}

\begin{figure*}
\includegraphics[width=8cm]{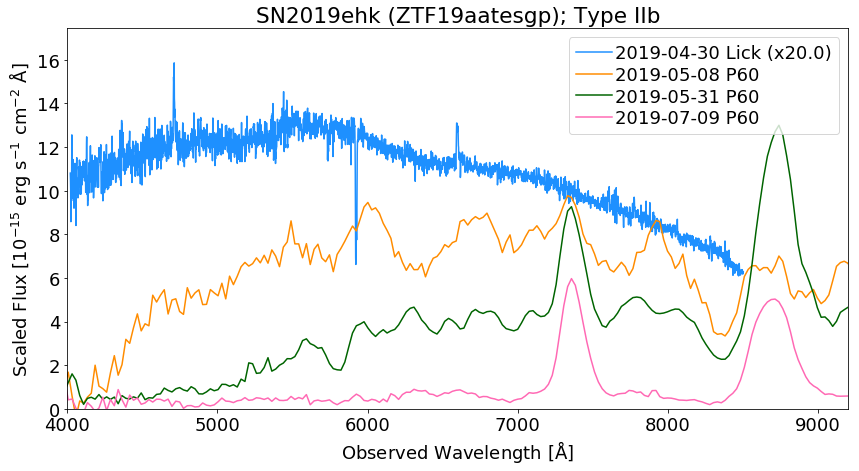}
\includegraphics[width=8cm]{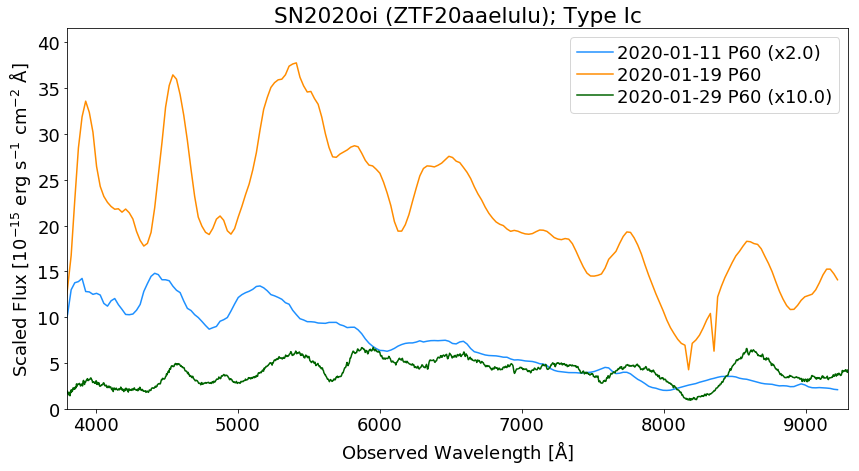}
\includegraphics[width=8cm]{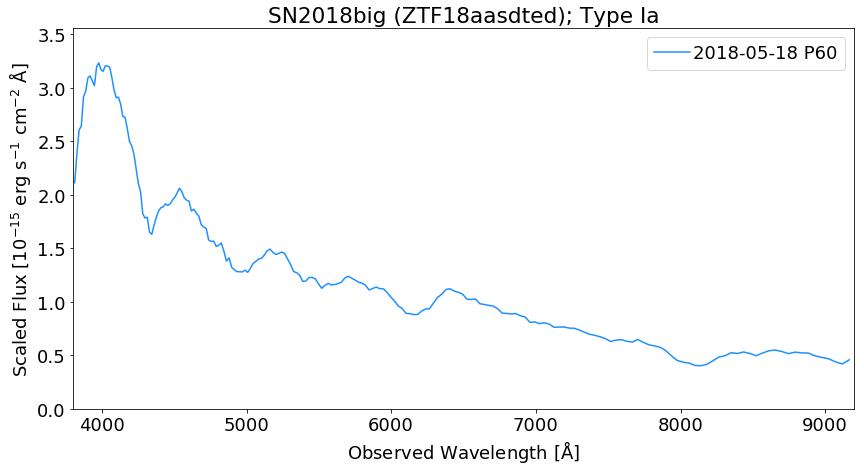}
\includegraphics[width=8cm]{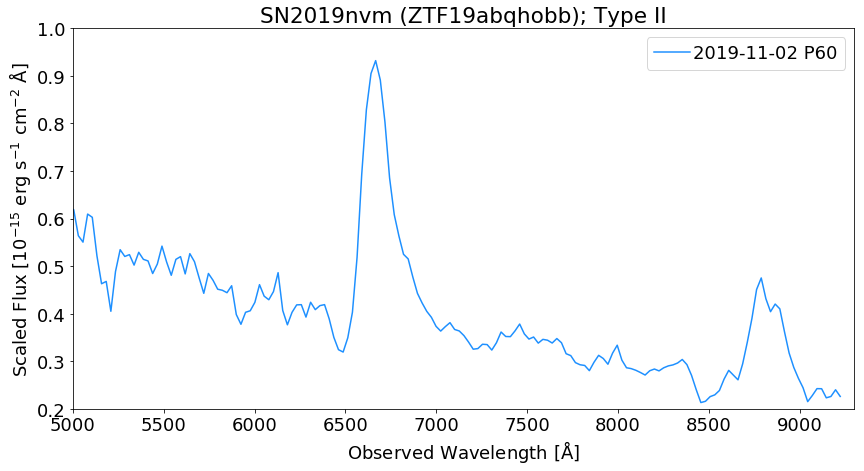}
\includegraphics[width=8cm]{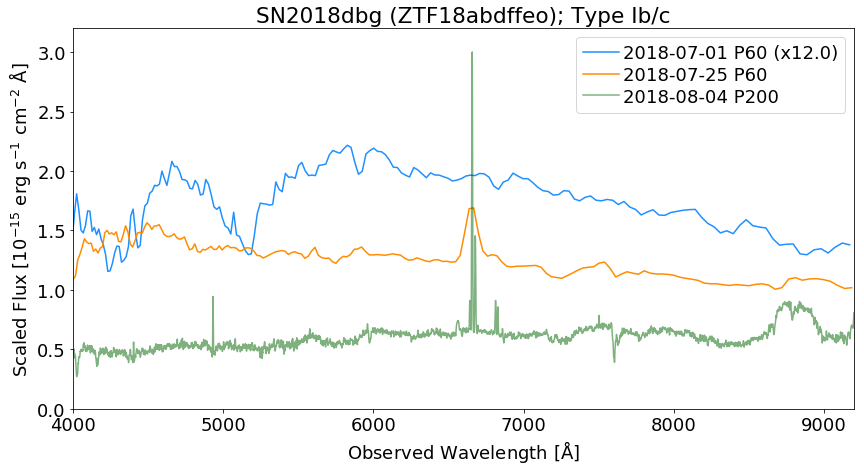}
\includegraphics[width=8cm]{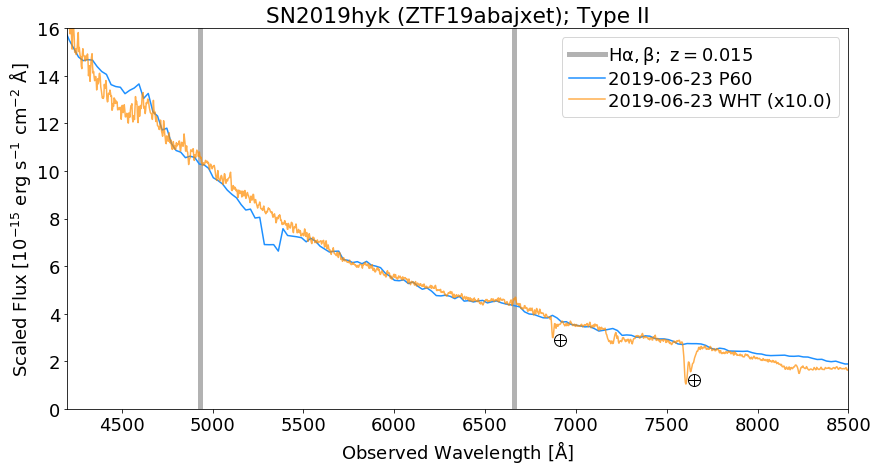}
\includegraphics[width=8cm]{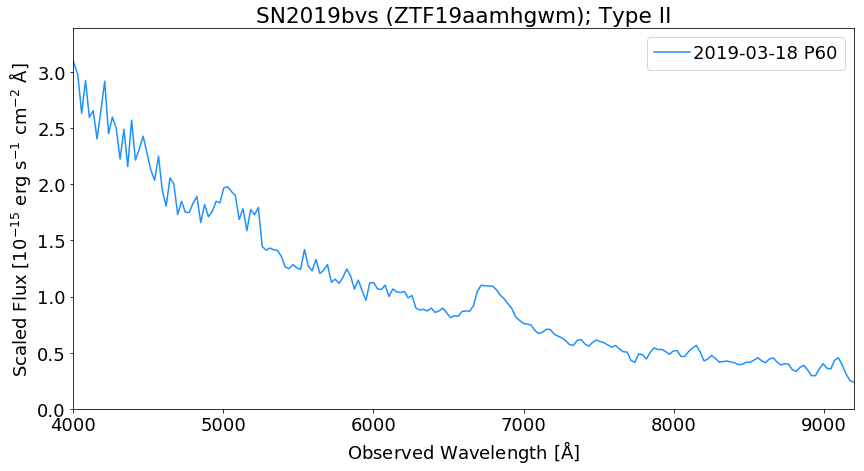}
\includegraphics[width=8cm]{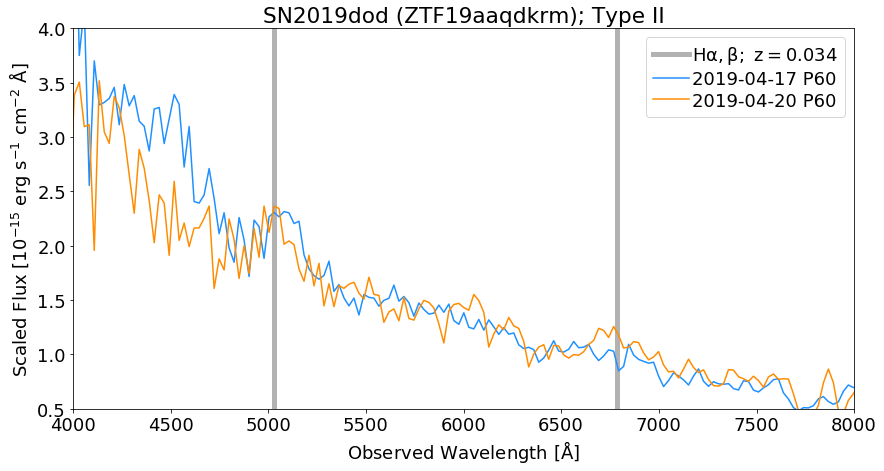}
\includegraphics[width=8cm]{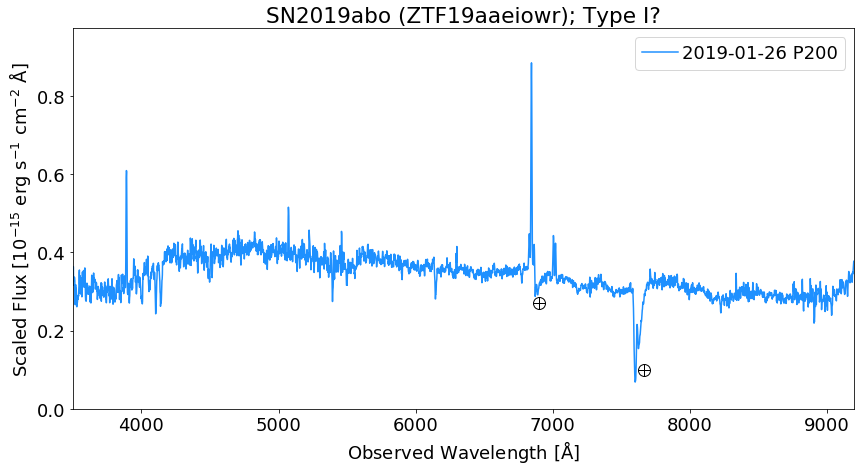}
\includegraphics[width=8cm]{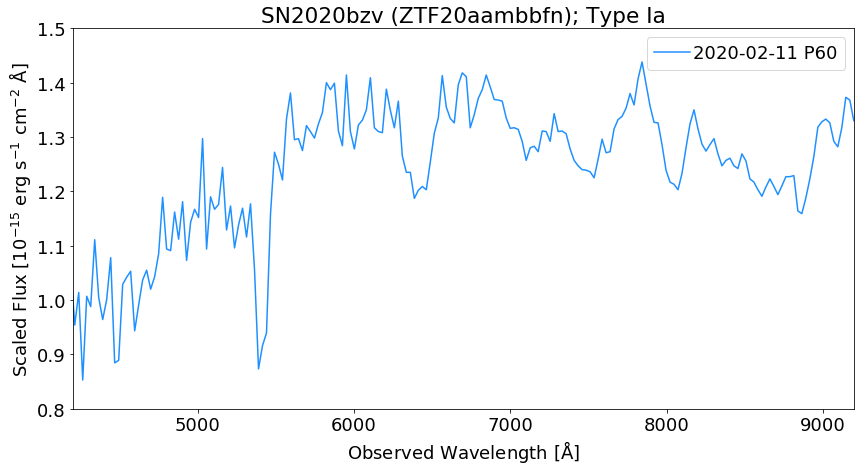}
\caption{Selected epochs of P60+SEDM (or P200) spectroscopy that are relevant to the classification of SN siblings with peak observed magnitude $\leq18.5$ mag in either the $g$ or $r$ filter (the Gold Sample). For SN\,2019ehk we also include the classification spectrum from Lick Observatory \citep{2019TNSCR.675....1D}. For SN\,2018dbg, the 2018-08-04 spectrum was published in \citet{2020arXiv200902347D}. For SN\,2019hyk, we also include the classification spectrum from the William Herschel Telescope \citep[WHT;][]{2019TNSCR1061....1F}.}
\label{fig:spec}
\end{figure*}

For these $5$ SN sibling pairs, Figure~\ref{fig:stamps} shows the location of each SN in the parent galaxy ($g$ band images from the PanSTARRS image cutout server; \citealt{2016arXiv161205560C,2020ApJS..251....3M}).
Figure~\ref{fig:phot} shows the ZTF public light curve data, and Figure~\ref{fig:spec} shows the BTS follow-up spectroscopy with the P60+SEDM which provided the spectroscopic classifications. 

The relevant discovery and classification information for each sibling pair of the Gold Sample, which motivates and supports the analysis in Section~\ref{sec:sibs}, is discussed below. 

\subsubsection{SN\,2019ehk (SN\,IIb)}

SN\,2019ehk (sibling to SN\,2020oi, below) was first discovered and reported to the Transient Name Server (TNS\footnote{\url{https://www.wis-tns.org/}}) on 2019 Apr 29 with a clear-filter brightness of $\sim16.5$ mag \citep{2019TNSTR.666....1G}.
Being so bright and in a nearby host galaxy Messier 100, SN\,2019ehk was detected by many other professional and amateur surveys, and spectroscopic monitoring began with its prompt classification as a SN\,Ib \citep{2019TNSCR.675....1D}.
However, as can be seen in the top-left panel of Figure~\ref{fig:spec} the spectrum of SN\,2019ehk evolved over time to resemble a SN\,IIb (H and He absorption) and then strong \ion{Ca}{II} features emerge.
Subsequent multi-wavelength follow-up out to late phases revealed SN\,2019ehk to be unlike typical "Ca-rich" transients associated with old stellar populations \citep{2012ApJ...755..161K}, and to likely be the core-collapse of a low-mass star exploding into a dense circumstellar material composed of its ultra-stripped envelope \citep[e.g.,][]{2020arXiv200502992N,2020ApJ...898..166J,2020arXiv200902347D}.

\subsubsection{SN\,2020oi (SN\,Ic)}

SN\,2020oi (sibling to SN\,2019ehk, above) was discovered by ZTF and first reported to the TNS by the ALeRCE broker on 2020 Jan 07 \citep{2020TNSTR..67....1F}, and classified as a Type Ic with an optical spectrum obtained within 2 days with the Goodman spectrograph at SOAR Observatory \citep{2020TNSCR..90....1S}.
The discovery of SN\,2020oi was more than ten days before peak $g$ band brightness, and rapid follow-up with Swift UVOT revealed a rising UV source \citep{2020TNSAN...8....1H}.
SN\,2020oi was also subsequently detected by optical imaging surveys ATLAS \citep{2018PASP..130f4505T}, the Young Supernova Experiment (YSE, using PanSTARRS1; \citealt{2020arXiv201009724J}), and Gaia Alerts \citep{2012gfss.conf...21W}, and monitored with optical spectroscopy \citep[e.g.,][]{2020ATel13404....1D}.
Optical photometric and spectroscopic follow-up was also obtained and reported by \citet{2020ATel13396....1P} and \citet{2020ATel13404....1D}.

Follow-up of SN\,2020oi in the radio in the days after discovery revealed a potential 10 GHz source and a confirmed 44 GHz source with the VLA \citep{2020TNSAN...9....1H,2020TNSAN..10....1H}, a detection at 15.5 GHz with AMI-LA \citep{2020TNSAN..11....1S}, and a detection at 5.1 GHz with e-MERLIN \citep{2020ATel13448....1M}.
\citet{2020ApJ...903..132H} presents the ZTF optical observations as well as radio data for SN\,2020oi; they model the radio observations and find that the density structure of the circumstellar material in the progenitor system might not follow the expected $r^{-2}$ distribution, but that otherwise SN\,2020oi is a standard Type Ic SN.
Optical and near-infrared observations of SN\,Ic 2020oi are also presented and analyzed by \citet{2020arXiv201000662R}, who confirm it to be a normal representative of the Type Ic class but which, uniquely, exhibits signatures of dust formation starting $>$60 days after explosion.

\subsubsection{SN\,2018big (SN\,Ia)}

SN\,2018big (sibling to SN\,2019nvm, below) was discovered on 2018 May 10 at 18.96 mag in the orange filter by the ATLAS survey \citep{2018TNSTR.609....1T}, and subsequently detected by the ZTF and PanSTARRS \citep{2016arXiv161205560C} surveys.
An optical spectrum with the P60+SEDM obtained on 2018 May 15 (shown in Figure~\ref{fig:spec}), was used to classify SN\,2018big as a Type Ia supernova \citep{2018TNSCR1071....1F}.

\subsubsection{SN\,2019nvm (SN\,IIP)}

SN\,2019nvm (sibling to SN\,2018big, above) was discovered on 2019 Aug 19 in ZTF imaging with an $r$ band brightness of 18.3 mag and reported by the AMPEL broker \citep{2019TNSTR1546....1N}, and subsequently detected by ATLAS and PanSTARRS.
As described by \citet{2019TNSCR1557....1H}, optical spectroscopy obtained with the FLOYDS instrument by the Las Cumbres Observatory Global SN Project on 2019 Aug 20 revealed a blue continuum and potential flash ionization features, classifying SN\,2019nvm as a Type II.
The spectrum obtained by ZTF on 2019 Nov 2 with the SEDM+P60 confirms this classification with its clear P-Cygni profile for H$\alpha$ (Figure~\ref{fig:spec}).
The ZTF light curve shown in Figure~\ref{fig:phot} exhibits the typical slow decline of Type IIP SNe.

\subsubsection{SN\,2018dbg (SN\,Ib/c)}

SN\,2018dbg (sibling to SN\,2019hyk, below) was discovered and first reported by ZTF on 2018 Jun 28 at $18.43$ mag in the $r$ band \citep{2018TNSTR.915....1F}, and subsequently detected and reported by Gaia Alerts on 2018 Jul 11.
An optical spectrum with the 5.1m Hale Telescope at Palomar Observatory on 2018 Aug 04 (green line, Figure~\ref{fig:spec}) was used to classify SN\,2018dbg as a Type Ib/c \citep{2018TNSCR1142....1F}.
\citet{2020ApJ...905...58D} describe how, once the host galaxy emission is subtracted from the 2018 Aug 04 spectrum, they could more clearly see photospheric-phase oxygen, calcium, and helium lines in order to classify SN\,2018dbg as a Ib-like event.
SN\,2018dbg does show a broad \ion{Ca}{II} triplet emission feature, but \citet{2020ApJ...905...58D} explain that nebular [\ion{Ca}{II}] is not seen and thus SN\,2018dbg is rejected from their sample of Ca-rich gap transients.
The light curve of SN\,2018dbg is poorly sampled but resembles a SN\,Ib/c in terms of its rise and fall, and has a $g-r$ color at peak brightness of $\sim$0.75 mag, which is consistent with SNe\,Ib/c as a population \citep{2015A&A...574A..60T}.
As quoted by \citet{2020ApJ...905...58D}, the peak intrinsic magnitude of SN\,2018dbg was $M_r=-16.6$ mag, which is on the faint side but by no means an outlier for the SN\,Ib/c class \citep{2014AJ....147..118R}.
All together, SN\,2018dbg appears to be a normal Type Ib/c supernova.

\subsubsection{SN\,2019hyk (SN\,IIP)}

SN\,2019hyk (sibling to SN\,2018dbg, above) was discovered by the All Sky Automated Survey for SuperNovae (ASAS-SN\footnote{\citet{2017PASP..129j4502K}; \url{http://www.astronomy.ohio-state.edu/asassn}}) on 2019 Jun 22 \citep{2019TNSTR1053....1S}, with a first detection magnitude of 17.1 in the SDSS-\textit{g} filter.
Classification of SN\,2019hyk as a Type II was reported by \citet{2019TNSCR1061....1F} using a spectrum from the ACAM instrument on the William Herschel Telescope, as shown in Figure~\ref{fig:spec} (orange line; the blue line shows a P60+SEDM spectrum from the same night, \citealt{2019TNSCR2826....1F}). 
\citet{2019TNSCR1061....1F} describe how the blue spectrum exhibits a weak, broad emission feature at H$\alpha$ consistent with SNe\,II, and an emission feature at $\sim$4580 \AA\ (observer-frame) consistent with the high-ionization signatures of shock breakout seen in young core-collapse SNe \citep[e.g.,][]{2014Natur.509..471G}.
The ZTF light curve of SN\,2019hyk in Figure~\ref{fig:phot} exhibits a $\sim$60-day decline of $\sim$0.5 mag in \textit{r} (and $\sim$1.2 mag in \textit{g}), which is consistent with a plateau and indicates SN\,2019hyk is a SN\,IIP  \citep[e.g.,][]{2012ApJ...756L..30A}.

\subsubsection{SN\,2019bvs (SN\,IIL)}

SN\,2019bvs (sibling to SN\,2019dod, below) was first discovered and reported by the ATLAS survey on 2019 Mar 16 at 18.47 mag in the orange-ATLAS filter \citep{2019TNSTR.385....1T}, and subsequently detected and reported by ZTF and the AMPEL broker, Gaia Alerts, and PanSTARRS.
An optical spectrum obtained with the P60+SEDM on 2019 Mar 18, shown in Figure~\ref{fig:spec}, showed the broad H$\alpha$ feature which classified SN\,201bvs as a Type II supernova \citep{2019TNSCR.411....1F}.
The light curve of SN\,2019bvs in Figure~\ref{fig:phot} reveals a relatively slower rise to peak and then a steady decline instead of a plateau, indicating that it is a Type IIL and not a IIP. 

\subsubsection{SN\,2019dod (SN\,IIP)}

SN\,2019dod (sibling to SN\,2019bvs, above) was first discovered and reported by ZTF on 2019 Apr 13 at $18.59$ mag in the $r$ band filter \citep{2019TNSTR.586....1F}, and subsequently detected and reported by ATLAS, MASTER \citep{2016RMxAC..48...42L}, Gaia Alerts, and PanSTARRS.
An optical spectrum from the SPRAT (SPectrograph for the Rapid Acquisition of Transients) at the Liverpool Telescope on 2019 Apr 16 revealed a blue continuum, similar to a young Type II SN \citep{2019TNSCR.589....1P}.
Subsequent spectra obtained with the P60+SEDM over the next several days, as shown in Figure~\ref{fig:spec}, reveal emergent broad hydrogen features that further suggest SN\,2019dod as a Type II. 
A plateau phase is clearly seen in the light curve of SN\,2019dod in Figure~\ref{fig:phot}, solidifying its classification as a Type IIP SN.

\subsubsection{SN\,2019abo (SN\,I?)}

AT\,2019abo (sibling to SN\,2020bzv, below) was discovered and first reported by ZTF on 2019 Jan 25 at $r{\sim}19.08$ mag \citep{2019TNSTR.141....1N}, and subsequently detected and reported by ATLAS.
No spectroscopic classification has been publicly reported for AT\,2019abo.
A P200+DBSP spectrum obtained 2019 Jan 26, as shown in Figure~\ref{fig:spec}, is dominated by emission from the host galaxy.
The light curve for AT\,2019abo shown in Figure~\ref{fig:phot} is sparsely sampled, but with a rise ($>$15 days) and fall ($\sim$1.25 mag in the first 15 days after peak) consistent with a Type I supernova (i.e., does not exhibit a slow decline or a plateau like Type II SN).
If it is a SN\,Ia, the peak brightness of $g{\sim}18.25$ mag suggests an extinction of about 1 mag, given a distance modulus of $\mu{\sim}36.4$ mag based on the host galaxy redshift of $z=0.043$ from the H$\alpha$ emission line.
If it a SN\,Ib/c, the brightest of which are $\sim$1 mag fainter than SNe\,Ia \citep{2011MNRAS.412.1441L}, the host-galaxy extinction could be minimal.
Either would be consistent with the SN's location in the disk of its inclined host galaxy (Figure \ref{fig:stamps}).
We note that the line-of-sight Milky Way extinction for the sky coordinates of this SN sibling pair is very low, $A_V \approx 0.02$ mag \citep{2011ApJ...737..103S}. 

\begin{figure}
\begin{center}
\includegraphics[width=8cm]{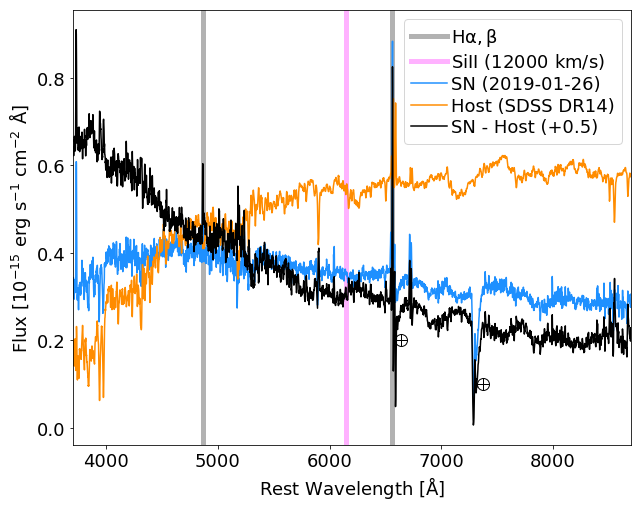}
\caption{The spectrum of AT\,2019abo from Fig. \ref{fig:spec} (blue), the host galaxy spectrum from SDSS DR14 (\citealt{2018ApJS..235...42A}; orange), and the host-subtracted SN spectrum (black). Grey and magenta lines mark the locations of hydrogen emission (host galaxy) and potential \ion{Si}{II} $\lambda$6355\AA\ absorption (a signature SN\,Ia feature), respectively. Earth symbols mark two strong atmospheric absorption features.}
\label{fig:SN2019abo_host_diff}
\end{center}
\end{figure}

In Figure~\ref{fig:SN2019abo_host_diff} we attempt to subtract host galaxy emission from the SN spectrum using an SDSS spectrum of the galaxy.
This is inappropriate because the SDSS spectrum includes emission from the host galaxy's core whereas the SN spectrum only includes emission at the SN location (and this host has a bright, compact nucleus).
The difference reveals a blue continuum which does not much resemble a SN\,Ia, and would be more similar to the massive CC events of SN\,Ib/c -- but no distinguishing features are revealed to confirm the type.
We thus refer to the type of SN\,2019abo as "\textit{SN~I?}" in Table~\ref{tab:sib_cands} to denote the uncertainty in its classification.

\subsubsection{SN\,2020bzv (SN\,Ia)}

SN\,2020bzv (sibling to SN\,2019abo, above) was first detected in ZTF images and reported by the ALeRCE broker on 2020 Feb 7 at $19.6$ mag in the $g$ filter \citep{2020TNSTR.413....1B}, and subsequently detected and reported by AMPEL and ATLAS.
An optical spectrum obtained on 2020 Feb 12 with the SPRAT at the Liverpool Telescope was used to classify SN\,2020bzv as a Type Ia supernova \citep{2020TNSCR.561....1P}.
The location of SN\,2020bzv was very close to (or projected on) the core of the host galaxy, and the light curve's peak brightness of $g{\sim}19.25$ mag suggests $\sim$2 mag of extinction; the light curve is also very clearly reddened.
A light curve fit for SN\,2020bzv using the SALT2 parameterization \citep{2007A&A...466...11G} in the {\tt SNCosmo} package \citep{SNCosmo} returns parameters $x_1 = -2.35 \pm 0.74$ ($\Delta m_{15} \approx 1.6$ mag) and $c=0.794 \pm 0.042$ (indicating significant reddening).
The spectrum of SN\,2020bzv in Figure~\ref{fig:spec} is also significantly reddened, but the \ion{Si}{II} $\lambda$6355 absorption feature clearly identifies the event as Type Ia.

\subsection{SN Siblings with Peak Brightness $<$19 mag}\label{ssec:sample_pLT19mag}

Table \ref{tab:sib_cands} lists three sibling SNe with peak brightnesses of $<$19 mag for which a secure classification was not possible: SNe 2019svq, 2019uej, and 2020hzk.
Although this paper's analysis does not use these SNe, future sibling SNe analyses might use them, and so here we provide a few further details about their potential classifications.

SN\,2019svq was detected by ZTF in only three epochs, all of them $r$ band, exhibiting a rise-time of at least $8$ days. 
Multiple epochs of spectroscopy were obtained with both the P60+SEDM and the P200 telescopes, but all spectra appear to be host-dominated.
The spectral type of SN\,2019svq remains unclear.

SN\,2019uej has a light curve with $8$ epochs over $\sim50$ days, $6$ of them during the transient's rise.
SN\,2019uej appears to have a color of $g{-}r$$\sim$$-0.2$ mag during the rise, and the light curve\footnote{A publicly viewable light curve for SN\,2019uej can be found at \url{https://lasair.roe.ac.uk/object/ZTF19acnwelq/}.} appears to be consistent with a SN\,Ia, like its sibling SN\,2019sik.
Two pre-max spectra were obtained of SN\,2019uej with the P60+SEDM; they are quite noisy but not inconsistent with a SN\,Ia at $z\sim0.10$ (i.e., they show potential signatures of \ion{Si}{II}, and not hydrogen).
Thus, it appears likely that SN\,2019uej was a Type Ia SN, the same as its sibling SN\,2019sik.
The spectra of sibling SN\,2019sik have a higher signal-to-noise ratio and indicate a redshift $z\sim0.1$ \citep{2019TNSCR2115....1D}, but using this redshift results in a peak absolute brightness that is overluminous by $0.5$--$0.8$ magnitudes (for both SNe, since they have very similar peak apparent brightnesses).

SN\,2020hzk has a fairly well-sampled ZTF light curve (11 epochs in 40 days)\footnote{The light curve of SN\,2020hzk is publicly viewable at \url{https://alerce.online/object/ZTF20aavpwxl}}.
It exhibits a rise time of $\sim15$ days; a color of $g-r\sim0$ mag until peak, after which it increases to $g-r\sim1$ mag by two weeks after peak; a decline rate of $\Delta g \sim 1.2$ mag during the 15 days after peak brightness; and an absolute peak brightness of $g\sim-19.05$ mag.
No spectra were obtained of this object, but its light curve's estimated peak intrinsic brightness ($\sim-19$ mag) and decline rate ($\sim1.3$ mag in the 15 days after peak) suggest that it is a Type Ia SN, the same as its sibling SN\,2019bbd.

\subsection{Additional ZTF Sibling SNe}\label{ssec:sample_additional}

There were eleven other confirmed or likely SN sibling pairs in the ZTF public survey (that we know of), and here we describe why they were not in the BTS sample.

The SN\,Ia sibling pairs ZTF20abatows/ZTF20abcawtk and ZTF19aakluwr/ZTF20acpqbue are not in our sample because in both cases, the latter SN of the two did not pass the BTS filter.
We furthermore find that relaxing the quality cuts imposed by \citet{2020ApJ...904...35P} results in seven additional SN sibling pairs:

\begin{itemize}
\item ZTF19abtuhqa (\emph{untyped}) / ZTF20abgaovd (SNIa); 
\item ZTF18ablqdru (\emph{untyped}) / ZTF20aattoch (SNII);
\item ZTF18aakaljn (SNIa-pec) / ZTF19acdtmwh (\emph{untyped});
\item ZTF19aalyecr (SNIa) / ZTF20aahkolj (\emph{untyped});
\item ZTF20aaeszsm (SNIa) / ZTF20abujoya (SNIa);
\item ZTF18adaadmh (SNIa) / ZTF20aamujvi (SNIa); and
\item ZTF19aatzlmw (SNIa) / ZTF20aaznsyq (\emph{untyped}).
\end{itemize}

Five of the above have at least one sibling without a classification (\emph{untyped}) due to lack of spectroscopy, and their light curve qualities are insufficient to compensate with a photometric classification.
Since they do not meet the quality cuts, and are not represented in the sample used to derive the completeness factors from \citet{2020ApJ...904...35P} which we use in Section \ref{sec:rates}, these SN sibling pairs are not used in this work.
However, future analyses which characterize the full BTS or ZTF sample would enable future ZTF SN siblings analyses to incorporate them.

\citet{Rahul_sibs} present two ZTF SN\,Ia siblings, SNe\,2019lcj and 2020aewj, in the same host galaxy at $z=0.0541$, within 0.6\arcsec\ of each other (and so both have the ZTF identifier ZTF19aambfxc).
SN\,2019lcj did not pass the BTS filter and so was not considered for this work.

Finally, \citet{2021arXiv210309937S} followed up on the curious case of ZTF20aamibse/AT\,2020caa, which appeared to exhibit a second outburst in 2021 of similar brightness.
They show that this second event is a SN\,Ia, offset by $\sim$1.3\arcsec\ from the original AT\,2020caa, and that the first event is also likely a SN (and not, e.g., a precursor outburst), and thus these two events were likely SN siblings as well.
However, the precursor event did not pass the BTS filter and so these two SNe were not included in our sample.

\section{SN Siblings and their Parent Galaxies}\label{sec:sibs}

Each of our sibling pairs and parent galaxies provide a unique opportunity to explore, e.g., the physics of SN progenitors, or their use as cosmological probes.
For the SN siblings in our Gold Sample (ZTF BTS events which reached a peak brightness of at least of $r{\sim}18.5$ mag), we provide a custom analysis of each sibling pair and host galaxy, in turn.

\subsection{SNe\,2019ehk \& 2020oi: A IIb--Ic Pair and their Older Siblings in Grand Design Spiral M100}\label{ssec:sibs_19ehk20oi}

The parent galaxy of Type IIb SN\,2019ehk and Type Ic SN\,2020oi is grand design spiral Messier 100 (M100), which is oriented face-on, as seen in the top-left panel of Figure~\ref{fig:stamps}.
Despite searching for past events in {\it all} of our Gold Sample parent galaxies, including SNe discovered by surveys other than ZTF, M100 is the only parent galaxy in our sample with more than two known sibling SNe. 
M100 has produced five other supernovae in the last 120 years: 1914A (untyped); 1901B, 1959E, and 1979C (SN\,I, SN\,I, and SN\,II, respectively; \citealt{1999A&AS..139..531B}); and 2006X (SN\,Ia; \citealt{2006IAUC.8667....1P,2006CBET..393....1Q}). 
The positions of all five previous SNe within M100 are shown along with ZTF BTS SNe\,2019ehk and 2020oi in Figure~\ref{fig:stamps}.

Our addition of stripped-envelope SNe 2019ehk and 2020oi to the M100 SN family allows for the unique opportunity to study the correlation between SN location and active star formation in a grand design spiral galaxy.
\citet{2016MNRAS.459.3130A} used a sample of 215 (non-sibling) SNe to show that, in grand design spirals like M100, SNe from higher-mass progenitors occur closer to the leading edges of spiral arms due to the shock-triggered star formation that occurs there (inside the corotation radius, the leading edges are the inner edges of spiral arms).
The corotation radius for M100 is ${\sim}10.5$ kpc \citep[e.g.,][]{2013MNRAS.428..625S}, which is equivalent to the image boundaries in the top-left panel of Figure~\ref{fig:stamps}.
Despite M100 being a prolific producer of SNe it is difficult to see (let alone confirm) this trend with its seven siblings, although SN\,II 1979C does appear on the outer edge of its spiral arm.
Despite SN\,Ia 2006X appearing near the inner edge of that same arm, it does not appear to be related to star formation activity (the bright clumps). 

The two ZTF BTS SNe in M100, SN\,IIb 2019ehk and SN\,Ic 2020oi, are the most centrally-located of the siblings (Figure~\ref{fig:stamps}) and also the two confirmed to be associated with massive stars (the rest being of Type I, Ia, or II). 
This agrees with previous large-sample analysis of (non-sibling) SNe-host offsets, which suggested that the regions of high stellar density are more efficient at forming the high stellar mass binaries that produce stripped-envelope SNe \citep[e.g.,][]{2012ApJ...759..107K,2014ApJ...789...23K}.
However, recent results from the Palomar Transient Factory show that the distribution of host offsets are very similar for all types of CC\,SNe \citep[e.g.,][their Fig. 12]{2020arXiv200805988S}.

\begin{figure}
\includegraphics[width=4.1cm]{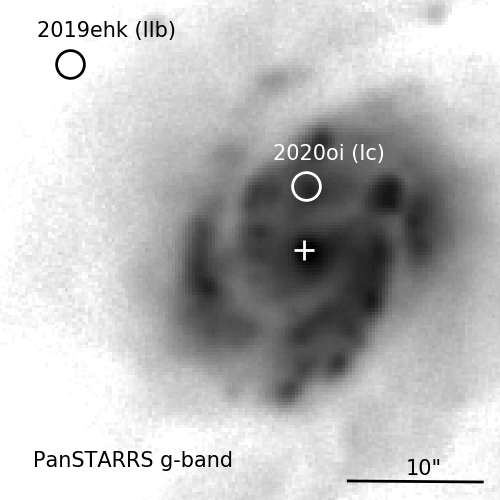}
\includegraphics[width=4.1cm]{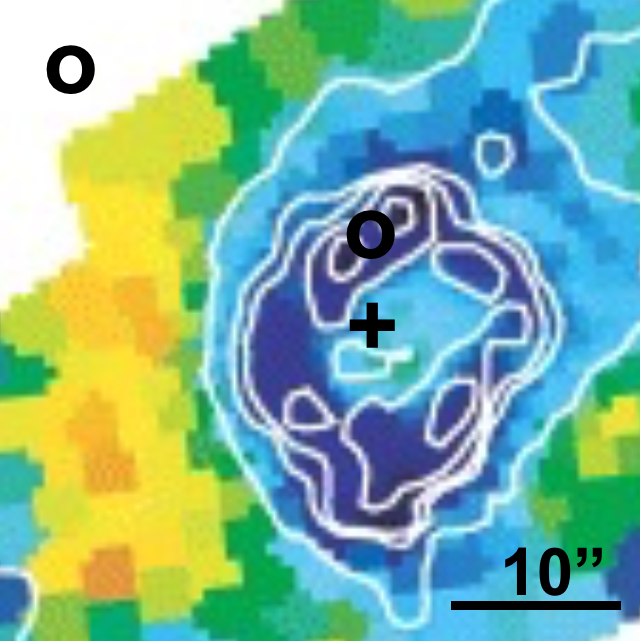}
\caption{{\it Left:} The central region of M100 in PanSTARRS $g$ band, with logarithmic scaling. {\it Right:} From Figure 7 of \citet{2006MNRAS.371.1087A}, the [\ion{O}{III}]/H$\beta$ emission line ratio (dark blue corresponds to a ratio of ${\sim}10^{-3}$, light blue to ${\sim}10^{-1}$) with H$\beta$ contours over-plotted in white. We have added the locations of SNe\,2020oi and 2019ehk. Lower ratios of the [\ion{O}{III}]/H$\beta$ emission line ratio indicate active star formation, and SN\,2020oi clearly originated in M100's circumnuclear ring of cool gas.}
\label{fig:M100}
\end{figure}

As M100 is a well-studied galaxy, we can use detailed maps of its star forming regions to analyze and compare the local environments of SN\,IIb 2019ehk and SN\,Ic 2020oi.
In the left panel of Figure~\ref{fig:M100} we show an enlargement of the central region of the optical PanSTARRS image from the top-left panel of Figure~\ref{fig:stamps}. 
The right panel of Figure~\ref{fig:M100} is from Figure 7 of \citet{2006MNRAS.371.1087A}.
It shows a map of the [\ion{O}{III}]/H$\beta$ ratio (with H$\beta$ contours), in which the dark blue regions indicate the sites of the most active star formation; we have added the locations of SN\,2020oi and 2019ehk to this star-formation map.
We can see that SN\,Ic 2020oi is clearly associated with the cool gas ring of star formation around the nucleus of M100.
Although SN\,IIb lies in a region without spectral coverage, the map suggests it lies in a region of lower star formation rate.
This agrees with other observations and models for CC\,SN progenitors in which SN\,Ic appear to be the explosions of the most massive and short-lived stars \citet[e.g.,][]{2012MNRAS.424.1372A}.

\subsection{SNe 2018big \& 2019nvm: Distance Estimates with a Ia--IIP Sibling Pair}\label{ssec:sibs_18big19nvm}

The parent galaxy of Type Ia SN\,2018big and Type II\,P SN\,2019nvm is UGC\,10858, a spiral oriented at high inclination, in which both SN\,2018big and 2019nvm appear to be embedded in the disk with a moderate offset from the host center (Figure~\ref{fig:stamps}).
The SDSS spectrum for UGC 10858 reveals signatures of star formation, and is the source of the redshift of $z=0.01815$ \citep{2020ApJS..249....3A}.
Type Ia and IIP SNe come from older and younger progenitor stars, respectively, and since both stellar populations are present in the disks of spiral galaxies it is not surprising to find both types of SNe in the same host.
Studies of the rates and properties of Type Ia SNe and their host galaxies have established that events associated with younger stellar populations tend to exhibit brighter, bluer, and broader optical light curves \citep[e.g.,][]{1995AJ....109....1H,2006ApJ...648..868S}.
Indeed, as discussed in Section~\ref{ssec:sample_gold}, SN\,2018big exhibited a slow decline rate (a broader light curve), a trait consistent with SN\,2018big being associated with a younger stellar population.

As discussed in Section~\ref{sec:intro}, Type Ia and IIP SNe can be used as cosmological distance estimators, and multiple SNe occurring in the same parent galaxy provide an opportunity to compare estimates in a scenario where the distance is known.
For this sibling pair of different SN types, we can compare distances estimated by two different methods. 

To Type Ia SN\,2018big we apply the SALT2 light curve fitter \citep{2007A&A...466...11G} to the ZTF $g$- and $r$-band photometry.
SALT2 returns light curve fit parameters of $x_1 = 0.72$, which translates to a light curve decline rate parameter of $\Delta m_{15}(B) = 0.98$ mag, which is at the low end of the SN\,Ia decline-rate distribution and indicates that SN\,2018big was also likely more luminous than average.
SALT2 returns light curve fit parameters of the apparent $B$-band magnitude, $m_B = 15.8$ mag, and the color excess, $c=0.11$.
These values agree also with those reported for SN\,2018big in \citet[][their Table 3]{2019ApJ...886..152Y}.

The SALT2 parameters are related to the distance modulus, $\mu$, via $\mu = m_B + \alpha*x_1 - \beta * c - M_B$, where $\alpha$ and $\beta$ are the width-luminosity and colour-luminosity relation coefficients and $M_B$ is the absolute $B$-band magnitude from an independent calibration of SNe\,Ia.
Using the $\alpha$ and $\beta$ relation coefficient from the ZTF Year 1 sample \citep{Dhawan_inprep}, we get a corrected peak apparent magnitude of 15.498 $\pm$ 0.04 mag.
We take the $M_B$ value of -19.326 $\pm 0.03$ mag from the tip of the red giant branch (TRGB) calibration of the SN~Ia luminosity \citep{2019ApJ...882...34F} to get a distance modulus of 34.824 $\pm$ 0.055 (statistical) mag.
Adding a systematic error of 0.15 mag for the intrinsic scatter of SN\,Ia we get $\mu = 34.824 \pm 0.16$ mag. 
We note that using a calibration of $M_B$ from Cepheid variables \citep{2019ApJ...876...85R} gives a $\mu$ value that is 0.1 mag lower.
In the following description of the distance for SN\,II-P 2019nvm, we assume an $H_0$ of 70 kms$^{-1}$Mpc$^{-1}$, which is what the TRGB calibration yields.
Hence, for the direct comparison of the SN\,II-P and SN\,Ia distance (see below), we use the value of $\mu$ derived using $M_B$ from the TRGB value.

For Type IIP SN\,2019nvm, \citet{Nance_19nvm} apply the SN\,IIP distance estimate method of \citet{2006ApJ...645..841N} to their time series of $g$, $r$, and $i$-band photometry and five optical spectra between 10 and 90 days after explosion.
They find an intrinsic $I$-band magnitude of $-17.84 \pm 0.14$ mag at 50 days, a distance modulus of $\mu = 34.95 \pm 0.26$ mag and a distance of $D = 97.6 \pm 12$ Mpc for SN\,2019nvm.
SN\,2019nvm will be included in a full cosmological analysis of ZTF SNe\,IIP in a forthcoming publication.

Finally, we can make a direct comparison of the two SN-derived distances for the parent galaxy: $\mu_{Ia} = 34.824 \pm 0.16$ mag and $\mu_{IIP} = 34.95 \pm 0.26$ mag.
These two values are discrepant by only $0.126$ mag ($<$1\%), and their combined errors are $0.30$ mag, so these particular distances have very good agreement within $0.4\sigma$.
Given that these two SNe are in the disk of an inclined galaxy, this could be interpreted as an encouraging sign that the distance estimates are not aversely affected by host dust -- at least in this one case.
However, analysis of larger samples (such as those discussed in Section~\ref{sec:intro}) remain necessary to robustly evaluate systematics in distance estimates for cosmology.

\subsection{SNe 2018dbg \& 2019hyk: Ib/c--IIP Siblings}\label{ssec:sibs_18dbg19hyk}

\begin{figure}
\begin{center}
\includegraphics[width=8cm]{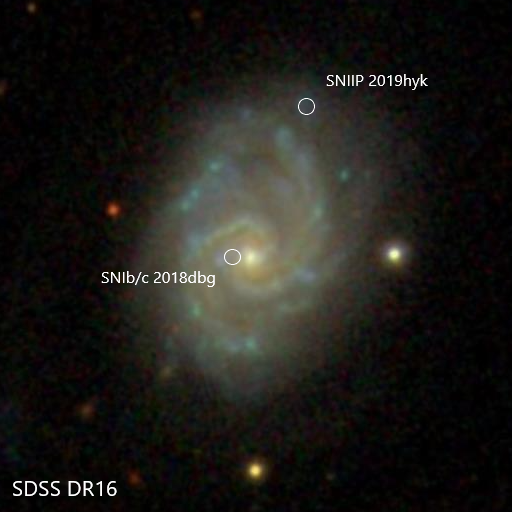}
\caption{An SDSS color image of the host galaxy IC4397 (north-up, east-left, field of view $\sim$2\arcmin), with the locations of SN\,Ib/c 2018dbg and SN\,IIP 2019hyk marked as open white circles.}
\label{fig:host_IC4397}
\end{center}
\end{figure}

The parent galaxy of Type Ib/c SN\,2018dbg and Type IIP SN\,2019hyk, IC\,4397, is a nearly face-on spiral galaxy (Figure \ref{fig:host_IC4397}, and member of the Coma Supercluster \citep{2010A&A...518A..10V}.
IC\,4397 has an observed redshift of $z=0.014737$ \citep{2002LEDA.........0P}, an apparent brightness of $\sim$13.04 mag in the SDSS $r$ band \citep{2005AJ....129.1755A}, and is classified as an AGN with an active \ion{H}{II} nucleus by \citet{2010A&A...518A..10V}.
Specifically, IC\,4397 is a Seyfert 2 galaxy, based on the infrared properties of its nucleus \citep{1987ApJ...321..233E,2001ApJ...557...39P,2007AJ....134.2006R}.
The SDSS spectrum reveals narrow H$\alpha$, [\ion{N}{II}] and [\ion{S}{II}], the classic signatures of star formation \citep[SDSS DR14,][]{2018ApJS..235...42A}. 

As discussed in Section~\ref{ssec:sibs_19ehk20oi}, the central regions of spiral galaxies have the younger stellar populations and/or higher stellar densities which could be more efficient at forming the high mass binary stars which are the progenitors of SNe\,Ib/c.
As expected, SN\,Ib/c 2018dbg is much closer to the core of the host than SN\,IIP 2019hyk, which is in the outskirts of the host.

\subsection{SNe 2019bvs \& 2019dod: A IIL-IIP Sibling Pair}\label{ssec:sibs_19bvs19dod}

\begin{figure}
\begin{center}
\includegraphics[width=7cm]{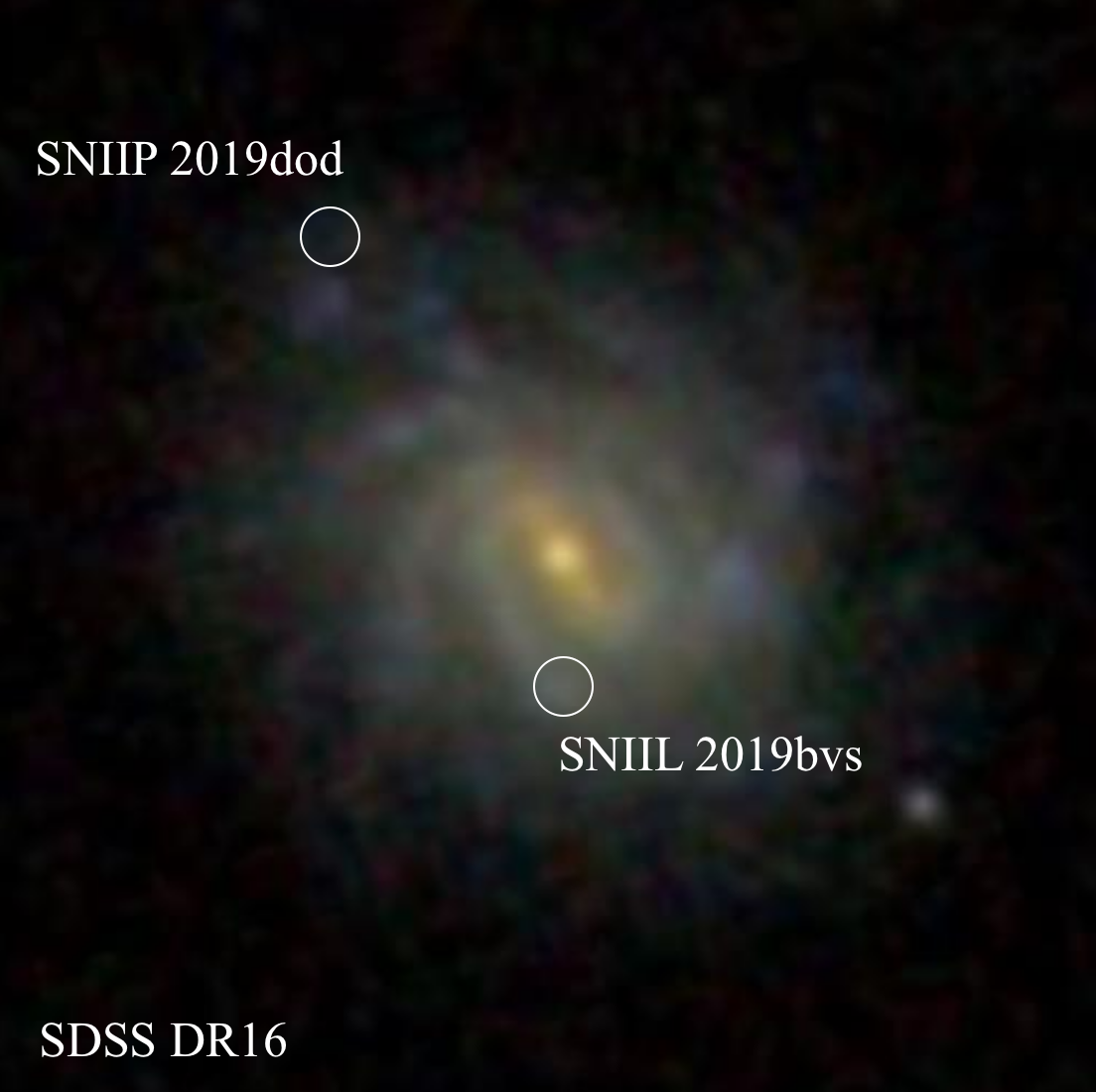}
\caption{An SDSS color image (north-up, east-left, field of view $\sim$1\arcmin) of the face-on barred spiral galaxy hosting SNe\,2019dod and 2019bvs.}
\label{fig:host_19dod19bvs}
\end{center}
\end{figure}

The unnamed parent galaxy of Type IIL SN\,2019bvs and Type IIP SN\,2019dod is a face-on barred spiral, as seen in the SDSS DR16 \citep{2020ApJS..249....3A} color image shown in Figure~\ref{fig:host_19dod19bvs}.
The color image exhibits clear evidence of star formation in its blue clumps (signatures of star formation are also seen in the SDSS spectrum, which we have not shown). 
This SN sibling pair provides the opportunity to compare the underlying stellar populations (and thus the progenitor scenarios) for the IIL and IIP subtypes of core collapse SNe.

It has been hypothesized that the difference between Type IIP, IIL, and IIb supernovae -- which exhibit light curves with a $\sim$100 day plateau, a slow decline ($\sim$1 mag in $\sim$60 days), and a fast decline ($\sim$1 mag in $\sim$20 days), respectively \citep{2012ApJ...756L..30A} -- could be the mass of the hydrogen envelope at the time of explosion, with less massive envelopes causing a more rapid decline.
This correlation would also indicate a trend with progenitor initial mass and age at the time of core-collapse, as more massive progenitors lose more of their hydrogen envelope and evolve towards collapse more rapidly.
Envelope mass as the underlying physical characteristic is supported by analyses of large samples of core collapse SN light curves which show that the Type IIP, IIL, and IIb are not distinct groups, but rather a continuum, as expected for a smooth distribution of envelope masses \citep{2014ApJ...786...67A,2014MNRAS.445..554F}.
Such a trend would result in Type IIP, IIL, and IIb SNe being found in regions with progressively younger stellar populations and more active star formation.
This trend was established by \citet{2012MNRAS.424.1372A}, who use H$\alpha$ emission as a tracer of active star formation in SN host galaxies. 
In particular, they show that the sites of Type IIL exhibit brighter H$\alpha$ emission than Type IIP SNe.

As is evident in Figure~\ref{fig:host_19dod19bvs}, Type IIL SN\,2019bvs is located in a bright blue knot in a spiral arm, whereas Type IIP SN\,2019bvs is in the host galaxy outskirts.
Thus, these two siblings exhibit the same trend in which Type IIL SNe are associated with younger stellar populations than Type IIP SNe.
As a final note, we remark that other Type IIL SN siblings in multi-SN parent galaxies do not show as clear a trend with active star formation, such as SN\,1926A in NGC 4303, SN\,1980K in NGC 6946, and SN\,1993G in Arp 299, all of which are in regions of fainter surface brightness and/or further from the galaxy's center than some of their SN\,IIP siblings \citep{2013A&A...550A..69A}.

\subsection{SNe 2019abo \& 2020bzv: Type I and Ia Siblings}\label{ssec:sibs_19abo20bzv}

The unnamed parent galaxy of Type I SN\,2019abo and Type Ia SN\,2020bvz is an inclined spiral galaxy with a compact nucleus and spectral signatures of a young stellar population (SDSS DR16; \citealt{2020ApJS..249....3A}).
Unfortunately, given the extreme reddening and extinction of SN\,2020bzv -- which is either in, or projected onto, the host galaxy's nucleus -- and the lack of a classification for SN\,2020abo, these sibling SNe cannot be used for any specific host-related science goals, but they are included in the relative rates analysis of Section~\ref{sec:rates}.

\section{SN Siblings Relative Rates}\label{sec:rates}

In this work we have identified 5 (10) pairs of SN siblings in the ZTF BTS sample of SNe with peak brightness $<$18.5 ($<$19.0) mag, which \citet{2020ApJ...904...35P} have shown to have a spectroscopic completeness of 93\% (75\%).
We refer to the 5 pairs with peak brightnesses $<$18.5 mag as the "Gold Sample", because they are part of the BTS sample with a spectroscopic completeness of 93\% and can thus be used for a relative rates analysis.
As described in Section~\ref{sec:sample}, our sample of BTS SNe extends about a month before and after the sample used in \citet{2020ApJ...904...35P}, but within that time-frame the completeness statistics still apply.
The number of SNe that we identify as siblings in the BTS sample and their breakdown by type is listed in Table~\ref{tab:SNnum}.
The ratios of BTS SNe by type, CC\,SN to SN\,Ia and SN\,Ib/c to SN\,II, are provided with 1$\sigma$ statistical errors from \citet{1986ApJ...303..336G} in Table \ref{tab:SNrat}.
For these relative rate ratios, we count SN\,2019abo as a Type Ia (although it might be a Type Ib/c) so that the ratio of CC:Ia and Ib/c:II are both \textit{lower limits}. 

For our analysis of the relative rates of SN siblings we start with a comparison to \citet{2013A&A...550A..69A}, who analyzed a sample of 2384 classified, hosted SNe from the Asiago catalog \citep{1999A&AS..139..531B} that were detected prior to 2012 May 23.
They found that 486 SNe in their sample shared a parent galaxy with one or more siblings: in other words, $\sim$20\% of their SNe had one or more siblings, and $\sim$10\% of their hosts had more than one SN.
Many of the SNe in the Asiago catalog come from galaxy-targeted surveys that prioritize the monitoring of galaxies that are most likely to host supernovae, whereas the ZTF is an all-sky non-targeted survey.
With ZTF we find only 10 (20) SNe with peak magnitudes $<$18.5 ($<$19) mag share parent galaxy, out of 1190 (1857) classified SNe total, or $\sim$0.8\% ($\sim$1\%).
This difference in the siblings percentage is  primarily due to the fact that the ZTF survey includes many more low-mass hosts which have proportionally lower supernova rates and are single parents.
In this comparison we must also consider that the Asiago catalog incorporates many detections from as early as 1885, and from a variety of surveys, and that it does not have an internally consistent detection and classification efficiency like the ZTF.

\begin{table}
\centering
\caption{The number of BTS SNe by type, for siblings and the full sample. SN\,2019abo is counted as a SN\,Ia here.}
\label{tab:SNnum}
\begin{tabular}{lcc}
\hline
Peak Brightness: & $<$18.5 mag & $<$19 mag   \\
Completeness: & 93\% & 75\% \\
\hline
\multicolumn{3}{l}{\textit{Siblings Sample}} \\
Ia      & 3   & 8   \\
CC      & 7   & 11  \\
II      & 5   & 8   \\
Ib/c     & 2   & 3   \\
\hline
\multicolumn{3}{l}{\textit{Full Sample}} \\
Ia      & 939   & 1454  \\
CC      & 312   & 495   \\
II      & 233   & 371   \\
Ib/c     & 79    & 124   \\
\hline
\end{tabular}
\end{table}

\begin{table}
\centering
\def\arraystretch{1.2}
\caption{Ratios of the number of BTS SNe by type, for siblings and the full sample, with statistical uncertainties derived from the binomial confidence limits of \protect{\citet{1986ApJ...303..336G}}.  SN\,2019abo is counted as a SN\,Ia, which makes both the CC:Ia and Ib/c:II ratios \textit{lower limits}.}
\label{tab:SNrat}
\begin{tabular}{lcc}
\hline
Peak Brightness: & $<$18.5 mag & $<$19 mag   \\
\hline
\multicolumn{3}{l}{\textit{Siblings Sample}} \\
CC:Ia  & $2.3$\raisebox{2pt}{$_{-1.4}^{+3.7}$} & $1.4$\raisebox{2pt}{$_{-0.7}^{+0.9}$}  \\
Ib/c:II & $0.4$\raisebox{2pt}{$_{-0.3}^{+0.8}$} & $0.4$\raisebox{2pt}{$_{-0.2}^{+0.5}$} \\
\hline
\multicolumn{3}{l}{\textit{Full Sample}} \\
CC:Ia  & $0.33 \pm 0.02$ & $0.34 \pm 0.02$ \\
Ib/c:II & $0.34 \pm 0.05$ & $0.33 \pm 0.04$ \\
\hline
\end{tabular}
\end{table}

\citet{2013A&A...550A..69A} found that core-collapse (CC) SN siblings were likely to be the same type: SN\,Ib/c with other SN\,Ib/c, and SN\,II with other SN\,II.
They explained that this match in sibling type indicates that stars form during bursts of $<10$ Myr, i.e., the delay time of a SN\,II, because continuous (non-bursty) star formation would result in the younger, more massive progenitor stars of Type Ib/c existing in the same regions as the less massive Type II progenitors.
Our sample of CC\,SN siblings does not clearly show this effect, as all three of our Type Ib/c have a Type II sibling, but this could simply be due to our relatively smaller sample size.
\citet{2013A&A...550A..69A} also demonstrate that their ratio of SN\,Ib/c to SN\,II SNe in multiple-SN hosts is $0.338 \pm 0.047$, higher than their Ib/c:II ratio in single-SN hosts, $0.274 \pm 0.021$.
This statistically significant increase in the Ib/c:II ratio in galaxies that host multiple SNe further illustrates how recent star formation and a young stellar population make a galaxy more likely to become a multiple-SN parent.  
We list the Ib/c:II ratio for our BTS siblings, $0.40_{-0.3}^{+0.8}$, in Table~\ref{tab:SNrat}, but find only a small, insignificant increase over the ratio for the full sample $0.34\pm0.05$.

Since SNe\,Ia are from older stellar populations they occur in both star-forming and elliptical galaxies \citep[e.g.,][]{2005ApJ...629L..85S}, however core-collapse SNe are almost never found in elliptical galaxies \cite[e.g.,][]{2012ApJ...753...68G,2013ApJ...769...39S,2019ApJ...887..127I,Irani_inprep}.
Although SNe\,Ia will outnumber CC\,SNe in a magnitude limited sample like ours, as previously discussed, CC\,SNe are more likely to appear in the same parent galaxy due to the bursty nature of star formation and their short delay times.
Thus, any multiple-SN parent galaxy sample will contain mostly star-forming galaxies, and the CC:Ia ratio will be much higher among siblings.
As shown in Table~\ref{tab:SNrat}, our CC:Ia ratio for the full BTS sample is $0.33\pm0.02$ and for the siblings sample is $2.3_{-1.4}^{+3.7}$, with the $<$18.5 mag limit.
If we expand the sample to include SNe that peaked brighter than $<$19 mag, we find the CC:Ia ratio among siblings drops to $1.4_{-0.7}^{+0.9}$.
The difference between the CC:Ia ratios for the peak $<$18.5 and $<$19 mag sibling samples seems large, but the two values are within their combined uncertainties.
\citet{2013A&A...550A..69A} find a ratio of CC:Ia events of $1.149\pm0.052$ in single-SN hosts, and a larger ratio of $1.946\pm0.186$ in multiple-SN host galaxies.
Although both surveys see the increase in the CC:Ia ratio in multiple-SN parent galaxies, the markedly lower CC:Ia ratio in the ZTF BTS full sample, $0.33$, compared to the CC:Ia ratio in the single-SN host sample of \citet{2013A&A...550A..69A}, $1.149$, is due to both the depth and wide-field nature of the ZTF survey. 

Since we have a magnitude-limited sample with high completeness, we can use the difference in the co-moving volumes for ZTF BTS SNe\,CC and SNe\,Ia to estimate the \emph{intrinsic} ratios.
Assuming average peak intrinsic brightness for Type Ia and II SNe are $-19$ and $-17$ mag respectively, and a flat cosmology with $H_0=70$ $\rm km\ s^{-1}\ Mpc^{-1}$ and $\Omega_M=0.3$, the co-moving volume within which they reach a peak apparent brightness of $<$18.5 mag are $0.110$ and $0.009$ $\rm Gpc^3$, respectively.
Thus, to convert the observed CC:Ia ratios in Table \ref{tab:SNrat} into intrinsic ratios, we multiple by a factor of $0.110/0.009 = 12.2$. 
Based on this we estimate an intrinsic CC:Ia ratio for siblings and the full sample to be approximately $\sim$28 and $\sim$4, respectively.

\section{Conclusions}\label{sec:conc}

The five ZTF sibling pairs presented in this work contribute to a growing amount of literature demonstrating the use of SNe -- and SN siblings in particular -- as cosmic lighthouses: signals of the characteristics of unresolved stellar populations and interstellar material in distant galaxies.
With our sample from $\sim$2 years of the ZTF public survey we have focused on the unique aspects of each family, and provided individual analyses of the siblings and their parent galaxies.
In general we found that, as expected, the SNe from more massive progenitor stars explode nearer the cores of their host galaxies, and/or in regions with more active star formation.
This small siblings sample cannot confirm (or reject) any SN progenitor models yet, but, these trends exemplify the kind of information that SN siblings bring to the broader discussion of progenitor populations
We have also provided the first comparative rates analysis of SN sibling rates in a complete population from an unbiased, well-characterized survey.
We find a lower ratio of CC\,SN to SN\,Ia than past surveys which targeted specific galaxies in order to maximize the number of SN detections -- and as the most common type of SN is CC, it is not surprising that past surveys found more CC\,SNe.

As ZTF continues over the next few years, the rate of discovery of ZTF SN siblings will continue to increase, and we will be able to shift our focus from individual families to analyses of the larger sample.
Based on the first two years of ZTF BTS SN siblings, for each new year of ZTF survey, 5 new siblings are identified \emph{per past year of survey} in the sample with peak brightness $<$18.5 mag. 
Thus at the end of year 2 we have 10 siblings; at the end of year 3 we expect another 10 new siblings for a total of 20; at the end of year 4 we expect 15 new siblings for a total of 35; and at the end of year 5 (fall 2023) we expect a total of 55 siblings.
In the larger, less complete sample with peak brightness $<$19 mag we can expect $>$100 ZTF BTS siblings by the end of 2023.
Looking forward to the Rubin Observatory and its 10-year long Legacy Survey of Space and Time (LSST; \citealt{2019ApJ...873..111I}), the detection of millions of SNe will bring the opportunity for significantly larger samples of SN siblings studies at greater cosmological distances and earlier epochs in the cosmic star formation history and the chemical evolution of the universe.

\section*{Acknowledgements}

Based on observations obtained with the Samuel Oschin Telescope 48-inch and the 60-inch Telescope at the Palomar Observatory as part of the Zwicky Transient Facility project. ZTF is supported by the National Science Foundation under Grant No. AST-1440341 and a collaboration including Caltech, IPAC, the Weizmann Institute for Science, the Oskar Klein Center at Stockholm University, the University of Maryland, the University of Washington, Deutsches Elektronen-Synchrotron and Humboldt University, Los Alamos National Laboratories, the TANGO Consortium of Taiwan, the University of Wisconsin at Milwaukee, and Lawrence Berkeley National Laboratories. Operations are conducted by COO, IPAC, and UW.

This work was supported by the GROWTH project funded by the National Science Foundation under Grant No 1545949.

SED Machine is based upon work supported by the National Science Foundation under Grant No. 1106171.

The PanSTARRS1 Surveys (PS1) and the PS1 public science archive have been made possible through contributions by the Institute for Astronomy, the University of Hawaii, the PanSTARRS Project Office, the Max-Planck Society and its participating institutes, the Max Planck Institute for Astronomy, Heidelberg and the Max Planck Institute for Extraterrestrial Physics, Garching, The Johns Hopkins University, Durham University, the University of Edinburgh, the Queen's University Belfast, the Harvard-Smithsonian Center for Astrophysics, the Las Cumbres Observatory Global Telescope Network Incorporated, the National Central University of Taiwan, the Space Telescope Science Institute, the National Aeronautics and Space Administration under Grant No. NNX08AR22G issued through the Planetary Science Division of the NASA Science Mission Directorate, the National Science Foundation Grant No. AST-1238877, the University of Maryland, Eotvos Lorand University (ELTE), the Los Alamos National Laboratory, and the Gordon and Betty Moore Foundation.

MLG acknowledges support from the DIRAC Institute in the Department of Astronomy at the University of Washington.
The DIRAC Institute is supported through generous gifts from the Charles and Lisa Simonyi Fund for Arts and Sciences, and the Washington Research Foundation.

MR has received funding from the European Research Council (ERC) under the European Union's Horizon 2020 research and innovation programme (grant agreement number 759194 - USNAC).

\section*{Data Availability}


The ZTF BTS photometry and classification spectra for all objects used in this work are publicly available via the TNS or alert brokers such as ANTARES, ALeRCE, and Lasair.
More information about the ZTF-I data release for the public survey are available via the ZTF website, \url{https://www.ztf.caltech.edu/}.



\bibliographystyle{mnras}
\bibliography{ms} 








\bsp	
\label{lastpage}
\end{document}